\documentclass[a4paper]{article}
\usepackage{epsfig,amssymb,amsmath}
\usepackage{multirow}
\usepackage[section]{placeins}
\def\half{\frac{1}{2}}

\begin{document}
\title{\vskip -70pt \begin{flushright} {\normalsize
DAMTP-2007-61} \\ \end{flushright} \vskip 80pt {\bf Light Nuclei as
Quantized Skyrmions} \vskip 20pt}
\author{Olga V. Manko\thanks{O.V.Manko@damtp.cam.ac.uk},\,\,\,\,
Nicholas S. Manton\thanks{N.S.Manton@damtp.cam.ac.uk}\,\,\,\,
and\,\,\,\,Stephen W. Wood\thanks{S.W.Wood@damtp.cam.ac.uk} \\ \\
DAMTP, Centre for Mathematical Sciences,\\University of Cambridge,
Wilberforce Road,\\ Cambridge CB3 0WA, UK} \maketitle
\begin{abstract}
We consider the rigid body quantization of Skyrmions with
topological charges 1 to 8, as approximated by the rational map
ansatz. Novel, general expressions for the elements of the inertia
tensors, in terms of the approximating rational map, are presented
and are used to determine the kinetic energy contribution to the
total energy of the ground and excited states of the quantized
Skyrmions. Our results are compared to the experimentally
determined energy levels of the corresponding nuclei, and the
energies and spins of a few as yet unobserved states are
predicted.
\end{abstract}

\section{Introduction}
The Skyrme model \cite{skyrme} is a nonlinear effective theory of
mesons, specifically pions. Its nonlinearity allows for the
existence of topological soliton solutions, labelled by an
integer-valued topological charge, $B$. A quantized Skyrmion of
topological charge $B$ is interpreted as a nucleus with baryon
number $B$.

The $B=1$ Skyrmion was first quantized by Adkins, Nappi and Witten
\cite{anw,an}. They provided the first calibration of the Skyrme
model, by fitting the model to the proton and delta masses. The
toroidal $B=2$ Skyrmion was quantized in \cite{kop,bc}, and the
energies corresponding to the ground state, representing the
deuteron, and excited states were calculated. This analysis was
extended in \cite{lms}, to allow the two single Skyrmions to
separate in the most attractive channel. This led to a more
accurate determination of the mean charge radius, as the deuteron
is rather loosely bound.

The interpretation of the nuclei helium-3 and hydrogen-3 (triton)
as quantized states of the $B=3$ Skyrmion was considered by Carson
\cite{Carson}, and the spins and energies were calculated. This
analysis was extended in \cite{Carson1} by a computation of the
static electroweak properties of the quantized Skyrmion.

In \cite{walhout}, the $B=4$ Skyrmion was semiclassically
quantized, and the ground state (corresponding to the alpha
particle) and first excited state were determined, and their
energies calculated. The results, though novel, involved
consideration of selected vibrational modes as well as rigid body
motion, and are difficult to generalize to higher baryon numbers.
Here we will consider the $B=4$ case from a different perspective,
which may easily be generalized to higher baryon numbers, and
enables us to compute the excitation energies of further excited
states.

Further results on the allowed spin and isospin states of
quantized Skyrmions for $B$ up to 8 and beyond have been obtained
by Irwin \cite{irwin}, and taken further by Krusch \cite{krusch}.
However, in this work, there were no estimates of the energies of
the states.

It is not easy to assess the qualitative success of the Skyrme
model just from the results for $B \le 4$. The nuclei have the
correct spin and isospin quantum numbers, but on the whole the
ground states represent nuclei which are too small and too tightly
bound. Nuclei with $B=2$ or $B=3$ have no excited states,
experimentally, so Skyrmion excited states based on rigid body
quantization are not meaningful, and one expects them to break up
into individual nucleons if further degrees of freedom are
included.

There have been a number of developments which make it worthwhile
to reassess these results on quantized Skyrmions, and it is also
possible to extend them to the range of baryon numbers $1 \le B
\le 8$. First, it has been noted that a reparametrization of the
Skyrme model is desirable to achieve a better fit to nuclear sizes
and related quantities like moments of inertia \cite{mantonwood}.
The Skyrme length scale should be roughly doubled, and
consequently the dimensionless pion mass parameter also doubled
(to keep the physical pion mass fixed). Doubling the pion mass
parameter has little effect on the qualitative character of
classical Skyrmion solutions up to $B=7$, but for $B \ge 8$ there
is a clear difference \cite{bs2}. The stable solutions are no
longer the hollow polyhedra found earlier for $B$ up to 22 and
beyond, but instead more dense structures closer to what one
expects for nuclei. In particular, for $B$ a multiple of four,
there are stable solutions which look like bound states of two or
more of the cubically symmetric $B=4$ Skyrmions \cite{bms}. We
shall analyse below the quantum states of the $B=8$ Skyrmion,
which is made up of two $B=4$ cubes, and compare with the states of
nuclei with $B=8$, including beryllium-8.

Another development is a better understanding of the quantization
rules for Skyrmions, the so-called Finkelstein-Rubinstein (FR)
constraints \cite{fr}, which encode the requirement that a
quantized $B=1$ Skyrmion is a spin $\half$ fermion. The FR
constraints combine the symmetry of a Skyrmion, for any value of
$B$, with the topology of the Skyrme model, to constrain the spins
and isospins of quantum states. Here the rational map ansatz comes
in \cite{houghton}. This gives a separation of variables between
the angular and radial dependence of the Skyrme field. True
solutions do not exactly exhibit this separation, but they do so
approximately. The ansatz gives a simple closed formula for the
angular dependence of a Skyrme field, and rotational symmetries
are easier to find than if one just has a numerical Skyrmion
solution. The optimised rational map ansatz gives good
approximations to true solutions up to $B=7$ (and far beyond for
zero or small pion mass). Even if it is a poor approximation, it
can still be helpful in the numerical search for true solutions,
and more importantly here, it is helpful in determining the effect
of the FR constraints. Krusch has recently found a simple formula
for determining the crucial signs that occur in the FR constraints
\cite{krusch}. This formula requires knowledge of the rational map
approximating the Skyrmion.

The rational map ansatz allows a further simplification, valid to
the extent that a Skyrmion is well approximated by the ansatz.
Kopeliovich noted that the moments of inertia (both rotational and
isorotational) of a Skyrme field described by the ansatz are
rather simpler than for a general Skyrme field \cite{kopel}, since
the effect of a rotation is just to rotate the map, leaving the
radial profile function invariant. We simplify Kopeliovich's
formulae further, taking advantage of the complex analytic character of
a rational map, and obtain formulae for the 36 components of the
spin/isospin inertia tensor. These can be accurately evaluated,
and it is easy to recognize if certain components vanish because
of symmetry. Using these moments of inertia, we estimate anew the
energies of ground and excited states of quantized Skyrmions over
the range of baryon numbers $1 \le B \le 4$, and for the first
time those in the range $5 \le B \le 8$. The quantization is based
on the established method of rigid-body quantization of rotations
and isospin rotations. Particularly interesting for us are the
states of the $B=6$ Skyrmion, since our reparametrization of the
Skyrme model \cite{mantonwood} was based on the mass and charge
radius of the lithium-6 nucleus. Also interesting are the
states for $B=8$, since the double cube $B=8$ Skyrmion has not
previously been quantized.

A problem for the Skyrme model that emerged in the work of Irwin
\cite{irwin}, is that the spin states of the $B=5$ and $B=7$
Skyrmions disagree with those of the corresponding nuclei in their
ground states. It has been suggested more than once (see e.g. \cite{olganick}) 
that it might be appropriate, for these baryon
numbers, to quantize a deformed Skyrmion with different symmetry.
This would make sense, especially if the allowed spins were
thereby reduced, making the spin energy smaller. The smaller spin
energy might more than compensate the increased classical energy
of the deformed Skyrmion. In this paper we are able to
quantitatively assess this idea. For $B=7$ it looks reasonable. A
ground state with the correct spin $\frac{3}{2}$ for the
lithium-7/beryllium-7 isodoublet can be obtained, and the
previously found spin $\frac{7}{2}$ state interpreted as the
observed, relatively low-lying second excited state. The spin
$\frac{1}{2}$ first excited state is still problematic, however.
For $B=5$ the situation is less satisfactory.

In the next section we review the Skyrme model, and briefly
describe the recent reparametrization of the Skyrme model using
the lithium-6 nucleus \cite{mantonwood}. Although this is very
important, we show that a reparametrization alone cannot solve all
the problems of the Skyrme model. In section 3 we describe the
rational map ansatz for Skyrmions. Section 4 deals with the
quantization of Skyrmions, which proceeds by parametrizing
time-dependent solutions through collective coordinates. Here, we
recall how the model is fermionically quantized by the imposition
of FR constraints. In section 5 we present expressions for the
inertia tensors which appear in the formula for the kinetic energy
operator, in terms of the approximating rational map. Sections 6
to 13 deal with the energy levels of quantized Skyrmions of baryon
numbers 1 to 8 respectively. In section 14 we provide a conclusion.

\section{The Skyrme Model}

The Skyrme model is defined in terms of an $SU(2)$-valued scalar,
the Skyrme field \cite{skyrme,manton}. It is a low energy
effective theory of QCD, becoming exact as the number of quark
colours becomes large \cite{wittenglobal,wittencurrent}. We call
the topological soliton solutions which it admits
\emph{Skyrmions}.

The Lagrangian density is given by
\begin{equation}
\label{eq:lag} \mathcal{L} =
\frac{F_\pi^2}{16}\,\hbox{Tr}\,\partial_\mu U
\partial^\mu U^{-1} + \frac{1}{32e^2}\,\hbox{Tr}\,[\partial_\mu U U^{-1},
\partial_\nu U U^{-1}][\partial^\mu U U^{-1},
\partial^\nu U U^{-1}] + \frac{1}{8} m_\pi ^2 F_\pi
^2\,\hbox{Tr}\,(U-1_2) \,,
\end{equation}
where $U(t,\mathbf{x})$ is the Skyrme field, $F_\pi$ is the pion
decay constant, $e$ is a dimensionless parameter and $m_\pi$ is
the pion mass.

Using energy and length units of $F_\pi / 4e$ and $2/eF_\pi$
respectively, we may express the Lagrangian as follows:
\begin{equation}
\label{eq:lagstandard} L=\int \left\{
-\frac{1}{2}\,\hbox{Tr}\,(R_\mu R^\mu) +
\frac{1}{16}\,\hbox{Tr}\,([R_\mu,R_\nu][R^\mu,R^\nu]) +
m^2\,\hbox{Tr}\,(U - 1_2) \right\} d^3 x \,,
\end{equation}
where we have introduced the $\mathfrak{su}(2)$-valued current
$R_\mu = (\partial_\mu U)U^{-1}$, and defined the dimensionless
pion mass parameter $m = 2m_\pi / eF_\pi$.

Field configurations of finite energy must satisfy
the boundary condition $U \rightarrow 1_2$ as
$|\mathbf{x}| \rightarrow \infty$. This compactifies $\mathbb R
^3$ to a 3-sphere of infinite size, and so topologically $U: S^3 \rightarrow S^3$ at a fixed
time. Field configurations $U$ therefore lie in topological
sectors labelled by their topological degree
\begin{equation}
B = \int B_0(\mathbf{x}) \, d^3 x\,,
\end{equation}
where
\begin{equation}
\label{eq:cur} B_\mu(x)=\frac{1}{24\pi^2}\,
\epsilon_{\mu\nu\alpha\beta}\,\hbox{Tr}\, \partial^\nu
UU^{-1}\partial^\alpha U U^{-1}\partial^\beta U U^{-1} \,.
\end{equation}
The degree $B$, which takes integer values, is identified with the
baryon number. We refer to $B_0$ as the baryon density.

The kinetic part of the Lagrangian $L$ is
\begin{equation}
\label{eq:kin} T = \int\left\{-\frac{1}{2}\,\hbox{Tr}\,(R_0R_0)-
\frac{1}{8}\,\hbox{Tr}\,([R_i,R_0][R_i,R_0])\right\} \, d^3x \,,
\end{equation}
and this is quadratic in the time derivative of the Skyrme field.
The rest of the Lagrangian (\ref{eq:lagstandard}) is (minus) the
potential energy:
\begin{equation}
E=\int\left\{-\frac{1}{2}\,\hbox{Tr}\,(R_iR_i)-
\frac{1}{16}\,\hbox{Tr}\,([R_i,R_j][R_i,R_j])-\,
m^2\hbox{Tr}(U-1_2)\right\}d^3x\,.
\end{equation}
Static Skyrmion solutions can be obtained by solving the
variational equations derived from $E$, or in practice by
numerically minimising $E$ in the sector with given $B$.

The parameters $e$ and $F_\pi$ can be fixed in a number of ways.
It has been common practice to use the set of parameters given in
\cite{an} with the physical pion mass taken into account,
specifically
\begin{equation}
\label{eq:prov} e=4.84,\,\,\,F_\pi = 108\,\hbox{MeV}\,\,\,
\hbox{and}\,\,\,m_\pi=138\,\hbox{MeV}\,\,\, (\hbox{which
implies}~m=0.528) \,.
\end{equation}
In \cite{an}, the values of $e$ and $F_\pi$ were tuned to
reproduce the masses of the proton and the delta resonance. This
parameter set was adjusted to optimise the predictions of the
model in the $B=1$ sector at the expense of the $B=0$ sector,
which requires $F_\pi = 186\,\hbox{MeV}$. It is not, therefore,
the optimal parameter set globally.

In \cite{mantonwood}, we proposed that in order for the Skyrme
model to more closely model nuclear properties, a
reparametrization would be desirable. We performed such a
reparametrization by matching the model in the $B=6$ sector with
properties of the lithium-6 nucleus, obtaining
\begin{equation}
\label{eq:newparam} e=3.26,\,\,\,F_\pi = 75.2\,\hbox{MeV}\,\,\,
\hbox{and}\,\,\,m_\pi=138\,\hbox{MeV}\,\,\, (\hbox{which
  implies}~m=1.125) \,.
\end{equation}

Figure 1 shows graphs of nuclear masses and static Skyrmion masses
per unit baryon number, using this new parameter set. The Skyrmion
quantum energies are not included. We observe that the graphs
intersect at $B=6$, as expected. It is clear that it is not
possible by a single parameter choice to correctly match nuclear
and Skyrmion masses for all baryon numbers and this remains the
case when the quantum spin and isospin energies are included. We
believe that calibrating the model in the $B=6$ sector is a
promising way to describe the properties of nuclei with $B \ge 4$.
For $B=1,2,3$ the Skyrmion energies are now too high, and neither
the nucleon mass nor delta resonance will be accurately fitted.

\begin{figure}[h!]
\begin{center}
\includegraphics[width=10cm]{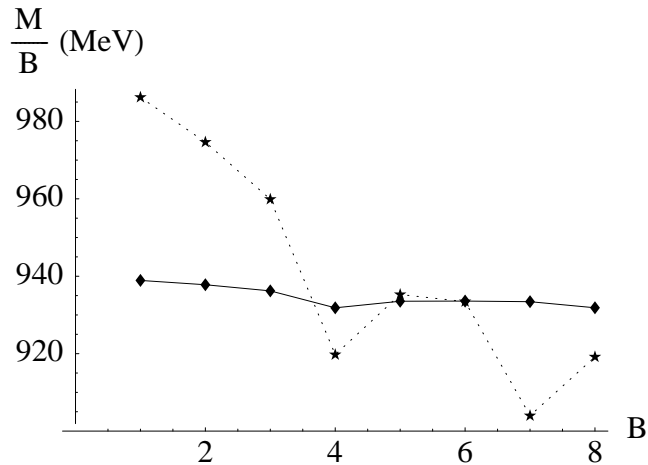}
\caption{Nuclear masses per unit baryon number $(M/B)$ (solid),
compared with static Skyrmion masses per unit baryon number (dotted).}
\end{center}
\end{figure}

For consistency, in the following sections we use the new
parameter set (\ref{eq:newparam}) throughout, making a few remarks
about the old parameters in Appendix B.

\section{The Rational Map Ansatz}
\label{sec:rma} We describe here the ansatz for Skyrme fields
which uses rational maps between Riemann spheres to describe their
angular behaviour \cite{houghton}. This has been shown to give
good approximations to several known Skyrmions, including all the
minimal-energy solutions up to $B=7$ (and much higher
$B$ when $m=0$). The rational maps have exactly the same
symmetries as the numerically known Skyrmions in almost all cases
($B=14$ is an exception \cite{bs1}).

Via stereographic projection, the complex coordinate $z$ encodes
the conventional polar coordinates as $z = \hbox{tan}(\theta
/2)e^{i\phi}$. Equivalently, the point $z$ on a sphere corresponds
to the radial unit vector
\begin{equation}
\label{eq:ster} \mathbf{n}_z = \frac{1}{{1 + |z|^2}}(z +
\bar{z},i(\bar{z}-z), 1 - | z|^2) \,,
\end{equation}
and inversely
\begin{equation}
z = \frac{(\mathbf{n}_z)_1 +
i(\mathbf{n}_z)_2}{1+(\mathbf{n}_z)_3}\,.
\end{equation}
The ansatz for the Skyrme field depends on a rational map $R(z) =
p(z) / q(z)$, where $p$ and $q$ are polynomials in $z$, and a
radial profile function $f(r)$. The target value $R$ is associated
with a point in the unit 2-sphere of the Lie algebra of $SU(2)$,
given by the unit vector
\begin{equation}
\label{eq:targetn} \mathbf{n}_R = \frac{1}{{1 + |R|^2}}(R +
\bar{R},i(\bar{R}-R), 1 - |R|^2) \,.
\end{equation}
The ansatz is then
\begin{equation}
\label{eq:ansatz} U(r,z) = \hbox{exp}\,(if(r)\mathbf{n}_{R(z)}
\cdot \boldsymbol{\tau}) \,,
\end{equation}
where $\tau_1$, $\tau_2$ and $\tau_3$ are Pauli matrices and $f(r)$ satisfies $f(0)=\pi$ and $f(\infty)=0$.

Using this ansatz, the baryon number is given by
\begin{equation}
B = \int \frac{-f'}{2\pi^2}{\left({\frac{\hbox{sin
}f}{r}}\right)}^2 \left( \frac{1 + |z|^2 }{ 1 + |R|^2}
\left\vert\frac{dR }{ dz }\right\vert
\right)^2\,\frac{2i\,dz\,d\bar{z}}{(1+|z|^2)^2}\,r^2\,dr \,,
\end{equation}
and it can be shown that this is an integer equal to the degree of
the rational map $R$.

The energy $E$ for a field of form (\ref{eq:ansatz}) is
\begin{equation}
\label{eq:e} E=4\pi \int_{0}^{\infty} \left(r^2 f'^2 + 2B\sin^2
f(f'^2+1)+ \mathcal{I}\,\frac{\sin^4 f}{r^2} + 2m^2r^2(1-\cos
f)\right) dr \,,
\end{equation}
in which $\mathcal{I}$ denotes the angular integral
\begin{equation}
\mathcal{I} = \frac{1}{4\pi} \int
\left(\frac{1+|z|^2}{1+|R|^2}\left\vert\frac{dR }{ dz }\right\vert
\right)^4\frac{2i\,dz\,d\bar{z}}{(1+|z|^2)^2}\,.
\end{equation}
To minimise $E$ one first minimises $\mathcal{I}$ over all maps of
degree $B$. The profile function $f(r)$ is then found by solving
the second order ODE that is the Euler-Lagrange equation for the
expression (\ref{eq:e}) with $B$ and $\mathcal{I}$ as fixed
parameters. Given the profile function, the energy is determined
by numerical integration. This gives the optimised rational map
ansatz, and we denote the minimised energy by ${\cal M}_B$. This
is our estimate for the true Skyrmion mass, for baryon number $B$. To obtain a physical value to
compare to a nuclear mass, one multiplies by the energy unit
$F_\pi / 4e = 5.76$\,MeV, obtaining a classical Skyrmion mass in MeV.

\section{Quantizing the Skyrme Model}
\label{sec:zero} Given a static Skyrmion $U_0(\mathbf{x})$, there
is generically a nine-parameter set of solutions, each with the
same energy, obtained by acting with the Euclidean group and
isorotations:
\begin{equation}
U(\mathbf{x})=A_1U_0(D(A_2)(\hbox{\bf x}-\mathbf{X}))A_1^{-1}
\end{equation}
where $A_1, A_2$ are $SU(2)$ matrices and $A_2$ is recast in the
$SO(3)$ form $D(A_2)_{ij} = \frac{1}{2} \hbox{Tr}(\tau_iA_2\tau_j
A_2^{-1})$. Semiclassical quantization is performed by promoting
the collective coordinates $A_1$, $A_2$, $\mathbf{X}$ to dynamical
degrees of freedom \cite{bc}. As we shall only be concerned with the
computation of spin and isospin, we shall ignore the translational
degrees of freedom $\mathbf{X}$ and quantize the solitons in their
zero-momentum frame.

Making the replacement $U(\mathbf{x}) \rightarrow
\hat{U}(\hbox{\bf x},t)=A_1(t)U_0(D(A_2(t))\hbox{\bf
x})A_1(t)^{-1}$, and inserting this into the Skyrme Lagrangian,
one obtains the kinetic contribution to the total energy
\begin{equation}
T=\frac{1}{2}a_i U_{ij} a_j - a_i W_{ij} b_j + \frac{1}{2}b_i
V_{ij} b_j \,,
\end{equation}
where
\begin{equation}
a_j = -i\,\hbox{Tr}\,\tau_j A_1^{-1}\dot{A}_1\,,\,\,\,b_j =
i\,\hbox{Tr}\,\tau_j \dot{A}_2 A_2^{-1} \,.
\end{equation}
$\mathbf{b}$ is the angular velocity in physical space, and
$\mathbf{a}$ is the angular velocity in isospace. The inertia
tensors $U_{ij}$, $V_{ij}$ and $W_{ij}$ are given by:
\begin{eqnarray}
\label{eq:u} U_{ij} &=& -\int \hbox{Tr}\, \left(T_iT_j +
\frac{1}{4}[R_k,T_i][R_k,T_j]\right)
\, d^3 x \,,\\
\label{eq:v} V_{ij} &=& -\int
\epsilon_{ilm}\,\epsilon_{jnp}\,x_lx_n\, \hbox{Tr}\,\left(R_mR_p +
\frac{1}{4}[R_k,R_m][R_k,R_p]\right) \, d^3 x \,,\\
\label{eq:w} W_{ij} &=& \int \epsilon_{jlm}\,x_l\,\hbox{Tr}\,
\left(T_iR_m + \frac{1}{4}[R_k,T_i][R_k,R_m]\right) \, d^3 x \,,
\end{eqnarray}
where $R_k = (\partial_k U_0)U_0^{-1}$ is the right invariant
$\mathfrak{su}(2)$ current defined previously and $T_i =
\frac{i}{2}\left[\tau_i,U_0\right]U_0^{-1}$ is also an
$\mathfrak{su}(2)$ current. The total energy, in terms of
collective coordinates, is just $T$ plus the constant ${\cal{M}}_B$,
the static mass of the Skyrmion.

We may write
\begin{equation}
T=\frac{1}{2} c^{\rm{T}} {\cal{W}} c\,,
\end{equation}
where $c^{\rm{T}} = (a_1,a_2,a_3,b_1,b_2,b_3)$, and the $6\times 6$
symmetric matrix ${\cal{W}}$ is given by
\begin{equation}
{\cal{W}} = \left(\begin{array}{cc} U & -W \\ -W^{\rm{T}} & V
\end{array}\right)\,.
\end{equation}

The momenta corresponding to $b_i$ and $a_i$ are the body-fixed
spin and isospin angular momenta $L_i$ and $K_i$ \cite{bc}:
\begin{eqnarray}
L_i &=& -W^{\rm{T}}_{ij}a_j + V_{ij}b_j\,, \\
K_i &=& U_{ij}a_j - W_{ij}b_j\,.
\end{eqnarray}
The usual space-fixed spin and isospin angular momenta $J_i$ and
$I_i$ are related to the body-fixed momenta by
\begin{equation}
J_i=-D(A_2)_{ij}^{\rm{T}} L_j,\,\,\,I_i = -D(A_1)_{ij}K_j \,.
\end{equation}
Defining $H^{\rm{T}} = (K_1,K_2,K_3,L_1,L_2,L_3)$, and using the relation
$H^{\rm{T}} = c^{\rm{T}} {\cal{W}}$, we find, provided $\det {\cal{W}} \neq 0$,
\begin{equation}
T= \frac{1}{2} H^{\rm{T}} {\cal{W}}^{-1} H\,.
\end{equation}

We now promote the four sets of classical momenta introduced above
to quantum operators, each individually satisfying the
$\mathfrak{su}(2)$ commutation relations. The Casimir
invariants satisfy $\mathbf{J}^2 = \mathbf{L}^2$ and $\mathbf{I}^2
= \mathbf{K}^2$.

The basic FR constraints, which apply to any Skyrmion, are that
physical quantum states $|\Psi\rangle$ should satisfy
\begin{equation}
e^{2\pi i\mathbf{n}\cdot\mathbf{L}}|\Psi\rangle = e^{2\pi
i\mathbf{n}\cdot\mathbf{K}}|\Psi\rangle = (-1)^B|\Psi\rangle \,,
\end{equation}
for any unit vector $\mathbf{n}$, which implies that for even $B$ the spin 
and isospin are integral,
and for odd $B$ they are half-integral. There are further
FR constraints on states if the Skyrmion has symmetries, and these
are simple to determine if the Skyrmion is described by the
rational map ansatz. A rational map, and hence the corresponding
Skyrmion, has a rotational symmetry if it satisfies an equation of the form
\begin{equation}
\label{eq:symm}R(M_2(z)) = M_1(R(z))\,,
\end{equation}
for some combination of $SU(2)$ M\"{o}bius transformations $M_2$
and $M_1$. $M_2$ corresponds to a rotation in physical space, and
$M_1$ to an isorotation. In general there will be a group ${\cal{S}}$
of such symmetries. We say that the map $R$ is ${\cal{S}}$-symmetric
if for each $M_2 \in {\cal{S}}$, there exists an $M_1$ such that
(\ref{eq:symm}) holds. For consistency, pairs $(M_2,M_1)$ must have the same
composition rule as in ${\cal{S}}$, so $R(M_2M'_2(z))=M_1M'_1(R(z))$.
The map $M_2 \rightarrow M_1$ is therefore a homomorphism. Note that it
is not possible to construct such a map from $M_1$ to $M_2$. This is
related to the fact that a Skyrmion may be invariant under a
rotation alone, but cannot be invariant under an isorotation alone. 

Consider a rotation in physical space by an angle $\theta_2$
about an axis $\mathbf{n}_2$, and an isorotation by an angle
$\theta_1$ about an axis $\mathbf{n}_1$. We recall that under such
a rotation, $z$ transforms to $M_2(z)$, given by \cite{krusch}:
\begin{equation}
M_2(z) = \frac{\left(\cos \frac{\theta_2}{2} +
i(\mathbf{n}_2)_3 \sin
\frac{\theta_2}{2}\right)z+\left((\mathbf{n}_2)_2
-i(\mathbf{n}_2)_1\right)\sin \frac{\theta_2}{2}}
{\left(-(\mathbf{n}_2)_2-i(\mathbf{n}_2)_1\right)\sin
  \frac{\theta_2}{2}z
+ \left(\cos \frac{\theta_2}{2} - i(\mathbf{n}_2)_3 \sin
\frac{\theta_2}{2}\right)}\,.
\end{equation}
Similarly, under such an isorotation, $R$ transforms to
$M_1(R)$, given by:
\begin{equation}
M_1(R) = \frac{\left(\cos \frac{\theta_1}{2} +
i(\mathbf{n}_1)_3 \sin
\frac{\theta_1}{2}\right)R+\left((\mathbf{n}_1)_2
-i(\mathbf{n}_1)_1\right)\sin \frac{\theta_1}{2}}
{\left(-(\mathbf{n}_1)_2-i(\mathbf{n}_1)_1\right)\sin
  \frac{\theta_1}{2}R + \left(\cos \frac{\theta_1}{2} - i(\mathbf{n}_1)_3
\sin \frac{\theta_1}{2}\right)}\,.
\end{equation}
So given a specific symmetry (\ref{eq:symm}) of a rational map, we use the above formulae to
determine the corresponding angles and axes of rotation and
isorotation. This is the data that is used in the conventional
formulae describing the effect of rotations on quantized rigid
bodies \cite{landau}.

For $\theta_2$ not an integer multiple of $2\pi$, $M_2$ only
leaves the points
\begin{equation}
z_{\mathbf{n}_2} = \frac{(\mathbf{n}_2)_1 + i(\mathbf{n}_2)_2}{1 +
(\mathbf{n}_2)_3}\,\,\,\hbox{and}\,\,\,z_{-\mathbf{n}_2} =
\frac{-(\mathbf{n}_2)_1 - i(\mathbf{n}_2)_2}{1 - (\mathbf{n}_2)_3}
\end{equation}
fixed. Similarly, $M_1$ only leaves $R_{\pm \mathbf{n}_1}$ fixed,
where $R_{\pm \mathbf{n}_1}$ are defined similarly. Therefore, for
the symmetry (\ref{eq:symm}) to hold, we have
\begin{equation}
R(z_{-\mathbf{n}_2}) = R_{\mathbf{n}_1} \,\,\,{\rm or}
\,\,\,R_{-\mathbf{n}_1} \,.
\end{equation}
Krusch showed that to correctly determine the FR constraint it is
important to choose the direction of the axis $\mathbf{n}_1$ so as
to satisfy the base point condition\footnote{We note that this
differs slightly from the condition given in \cite{krusch}. This
is because we are working with the inverse of the isospatial
M\"{o}bius transformation which was considered there.}
\begin{equation}
\label{eq:base} R(z_{-\mathbf{n}_2}) = R_{-\mathbf{n}_1} \,.
\end{equation}
The symmetry (\ref{eq:symm}) then leads to the following FR
constraint on the wavefunction:
\begin{equation}
\label{eq:freqn} e^{i\theta_2\mathbf{n}_2\cdot\mathbf{L}}
e^{i\theta_1\mathbf{n}_1\cdot\mathbf{K}}|\Psi\rangle =
\chi_{FR}|\Psi\rangle \,,
\end{equation}
a representation-independent statement, in which $\mathbf{L}$ and
$\mathbf{K}$ are the body-fixed spin and isospin operators
respectively, and the FR sign $\chi_{FR} = \pm 1$. The combined
rotation and isorotation corresponding to $M_2$ and $M_1$
respectively at most change the state by a sign
factor, $\chi_{FR}$. It was proved in \cite{krusch} that the value
of $\chi_{FR}$ for a given symmetry of a rational map only depends
on $\theta_2$ and $\theta_1$, where the angles have unambiguous signs
because of the base point condition (\ref{eq:base}), and is given by
\begin{equation}
\chi_{FR} = (-1)^{\cal{N}} \,,\,\,\,\hbox{where}
\,\,\,{\cal{N}}=\frac{B}{2\pi}(B\theta_2 - \theta_1) \,.
\end{equation}
The FR signs $\chi_{FR}$ form a 1-dimensional representation of
the symmetry group ${\cal{S}}$ of the Skyrmion.

A basis for the wavefunctions is given by $|J,L_3\rangle \otimes |
I,K_3\rangle$, the tensor product of states of a rigid body in space and
a rigid body in isospace. Here we suppress the additional labels $J_3$ and
$I_3$ which can take any values in the usual ranges allowed by $J$
and $I$. $J_3$ is the physically meaningful projection of spin on
the third space axis, and $I_3$ is the conventional third
component of isospin. When we come to consider specific rational
maps and the associated FR constraints, we seek low-energy states
which are allowed by the FR constraints, and may represent the
spin and isospin operators appearing on the left-hand side of
(\ref{eq:freqn}) by Wigner $D$-matrices, acting on the $(2J+1)
\times (2I+1)$-dimensional space of wavefunctions.

Another advantage of the rational map ansatz 
is that it clearly illustrates any reflection symmetries of the Skyrmion,
which enables one to determine the effect of the inversion operator $\cal{P}$, which
acts as
\begin{equation}
{\cal{P}}: U(\mathbf{x}) \rightarrow U^{\dag}(-\mathbf{x})\,,
\end{equation}
and is baryon number preserving.
For rational maps, the inversion $\mathbf{x} \rightarrow -\mathbf{x}$ corresponds to $z \rightarrow -1/\bar{z}$, and the inversion $U \rightarrow U^{\dag}$ corresponds to 
$R \rightarrow -1/\bar{R}$. A rational map, and hence the corresponding Skyrmion, has a reflection symmetry
if it satisfies an equation of the form
\begin{equation}
-1/\overline{R(M_2(z))} = M_1(R(-1/\overline{z}))\,.
\end{equation}
In this case, ${\cal{P}}$ is equivalently given by the combination of rotation and isorotation corresponding
to $M_2$ and $M_1$ occurring here. The parity of a quantum state is then the eigenvalue of the state
when acted upon by the operator $e^{i\theta_2\mathbf{n}_2\cdot\mathbf{L}}
e^{i\theta_1\mathbf{n}_1\cdot\mathbf{K}}$ derived from $M_2$ and $M_1$. There is, however, an ambiguity in the definition 
of ${\cal{P}}$ which was first explained in \cite{irwin}. Given a candidate parity operator ${\cal{P}}_0$ for a given Skyrmion, we can also represent the operator by
${\cal{P}}_0$ times any element of the symmetry group of the classical solution.
If the FR sign of a particular symmetry element is $-1$, then these 
two choices for $\cal{P}$ give different results. In particular, there is this problem for odd $B$: given a parity operator ${\cal{P}}_0$, we can also 
represent the operator by ${\cal{P}}_0 e^{2\pi i \mathbf{n}\cdot\mathbf{L}}$,
where $\mathbf{n}$ is any unit vector. As $2\pi$ rotations have associated FR
signs 
of $-1$ for odd $B$, we see that these two choices differ when acting on states. For each of the cases $B=1$ to 8, we make particular choices for the parity operators, 
which we believe to be the most natural. We note that despite the ambiguity in the definition
of $\cal{P}$ for a given Skyrmion, the relative parities of the Skyrmion's quantum states are fixed.

\section{Tensors of Inertia for Rational Map Skyrmions}

Kopeliovich \cite{kopel} first presented general formulae for the
inertia tensors of rational map Skyrmions. Writing $U_0 =
\exp(if(r)\mathbf{n}\cdot\boldsymbol{\tau})$, these can be
expressed as follows \cite{kopel}:
\begin{equation}
U_{ij} = 2\int \sin^2 f \left[(\delta_{ij}-n_i n_j)(1+f'^2) +
\sin^2 f\,
\partial_k n_i\partial_k n_j\right] d^3x\,,
\end{equation}
\begin{equation}
V_{ij} = 2\int \sin^2 f \Big[\left(1+f'^2+\sin^2 f\,\partial_k
  n_s\partial_k n_s\right)\left(\partial_m n_r\partial_m
n_r(r^2\delta_{ij}-x_i x_j)-\partial_i n_r
\partial_j n_r \,r^2\right)
\end{equation}
\begin{equation*}
-\sin^2 f \left(\partial_m n_s\partial_k n_s\partial_m
n_r\partial_k n_r(r^2\delta_{ij}-x_i x_j)-r^2\partial_i
n_r\partial_k n_r\partial_j n_s\partial_k n_s\right)\Big] d^3 x\,,
\end{equation*}
\begin{equation}
W_{ij} = 2 \int \epsilon_{jlm} \epsilon_{isp}x_l n_s \sin^2 f
\big[ \left(1+f'^2)\,\partial_m n_p +\sin^2 f \,\partial_k n_r
(\partial_k n_r \partial_m n_p - \partial_m n_r \partial_k
n_p\right)\big]\,d^3 x\,.
\end{equation}
These formulae for the inertia tensors assume that $f$ depends only on
$r$, and $\mathbf{n}$ depends only on the angular coordinates
$\theta$, $\phi$; further simplifications can
be made if we assume that $\mathbf{n}$ depends just on a rational
function $R(z)$ as in (\ref{eq:targetn}). In order to obtain these
simplified formulae, we find it helpful to write the ${\mathbb{R}}^3$
metric and volume element in terms of $r$, $z$ and $\bar{z}$:
\begin{equation}
ds^2 = dr^2 + r^2d\theta^2 + r^2\sin^2 \theta \, d\phi^2 = dr^2 +
\frac{4r^2\,dz\,d\bar{z}}{(1+|z|^2)^2} = g_{\alpha \beta}dx^\alpha
dx^\beta\,,
\end{equation}
\begin{equation}
d^3x = \frac{4r^2\,dr\,dz\,d\bar{z}}{(1+|z|^2)^2}\,,
\end{equation}
and to replace Cartesian derivatives with derivatives with respect
to $r$, $z$ and $\bar{z}$. The products of commutators in
(\ref{eq:u},\ref{eq:v},\ref{eq:w}) may then be rewritten in these
coordinates:
\begin{equation}
[R_k,\cdots ][R_k,\cdots ]=g^{rr}[R_r,\cdots ][R_r,\cdots ] +
g^{z{\bar{z}}}[R_z,\cdots ][R_{\bar{z}},\cdots ] +
g^{{\bar{z}}z}[R_{\bar{z}},\cdots ][R_z,\cdots ]\,,
\end{equation}
where $R_z = (\partial_z U_0)U_0^{-1}$ etc. We also have
\begin{equation}
-i\epsilon_{jlm}x_lR_m = (l_jU_0)U_0^{-1} = \mu_{j}R_z -
\bar{\mu}_{j}R_{\bar{z}}\,,
\end{equation}
where
\begin{eqnarray}
l_1 &=& -\frac{1}{2}\left((1-z^2)\frac{\partial}{\partial z}
-(1-{\bar{z}}^2)\frac{\partial}{\partial {\bar{z}}}\right)\,, \\
l_2 &=& -\frac{i}{2}\left((1+z^2)\frac{\partial}{\partial z}
+(1+{\bar{z}}^2)\frac{\partial}{\partial {\bar{z}}}\right)\,, \\
l_3 &=& z\frac{\partial}{\partial z} - {\bar{z}}\frac{\partial}
{\partial {\bar{z}}}\,,
\end{eqnarray}
so
\begin{equation}
\mu_{j} = \left( -\frac{1}{2}(1-z^2),\,
-\frac{i}{2}(1+z^2),\,z\right)\,.
\end{equation}

We ultimately find that for the rational map ansatz, the tensors
of inertia $U_{ij}$, $V_{ij}$ and $W_{ij}$ can be expressed in the
following form:
\begin{equation}
\Sigma_{ij} = 2 \int \sin^2 f
\,\frac{C_{\Sigma_{ij}}}{(1+|R|^2)^2} \left(1+f'^2+\frac{\sin^2
f}{r^2} \left(\frac{1+|z|^2}{1+|R|^2}\left| \frac{dR}{dz}\right|
\right)^2 \right) d^3 x\,,
\end{equation}
where $\Sigma = (U,V,W)$ and the quantities $C_{\Sigma_{ij}}$
(which are given explicitly in Appendix A) are functions of
the variables $z$ and $\bar{z}$ only. In what follows, we use the
above formula to numerically determine the elements of the inertia tensors, for
a given rational map and profile function. The numerical values we
obtain are, of course, in Skyrme units. To convert to physical
values we must multiply these by the mass scale and by the square
of the length scale: $(F_\pi / 4e) \times (2/eF_\pi)^2 = 1/e^3
F_\pi$, obtaining quantities in inverse MeV. With the new parameter set (\ref{eq:newparam}), 
$e^3 F_\pi = 2613\hbox{\,MeV}$. Although our optimal rational
maps are familiar \cite{houghton}, the profile functions have all
been calculated anew using a shooting method\footnote{due to
Bernard M. A. G. Piette, University of Durham, UK}. In Fig. 2 we
plot the profile functions for $B = 1$ to 8, using the new
dimensionless pion mass parameter $m = 1.125$.
\begin{figure}[h!]
\begin{center}
\includegraphics[width=11cm]{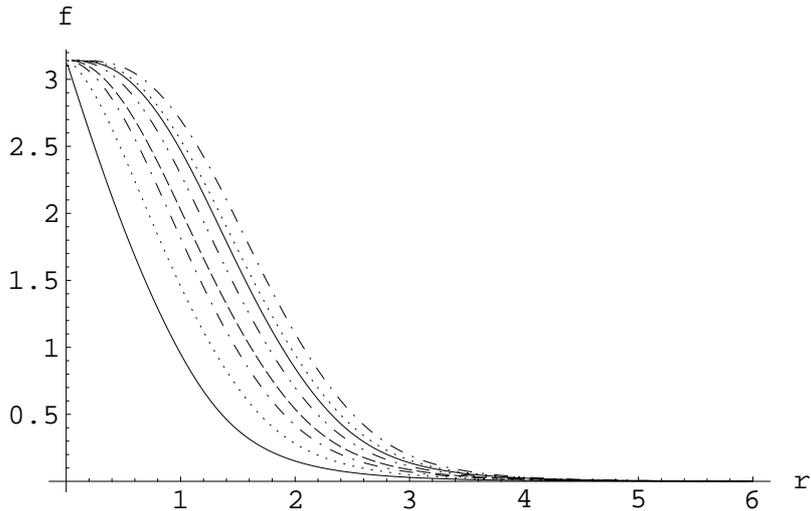}
\label{fig:profiles} \caption{The profile functions $f(r)$ for
$B=1$ to 8. $B$ increases from left to right.}
\end{center}
\end{figure}

\section{$B=1$}
The rational map describing the single Skyrmion is given by
$R(z)=z$, which is $O(3)$ symmetric. We find that the inertia
tensors are each proportional to the unit matrix, satisfying
$U_{ij}=V_{ij}=W_{ij}=\lambda \delta_{ij}$, where $\lambda$ is
given by
\begin{equation}
\lambda = \frac{16\pi}{3} \int r^2 \sin^2 f
\left(1+f'^2+\frac{\sin^2 f}{r^2} \right) dr\,.
\end{equation}
Numerically we compute $\lambda =45.1$.

The FR constraints associated with spherical symmetry are
\begin{equation}
e^{i \theta {\mathbf{n}} \cdot {\mathbf{L}}}e^{i \theta {\mathbf{n}}
\cdot {\mathbf{K}}} |\Psi\rangle = |\Psi\rangle\,,
\end{equation}
where $\theta$ and $\mathbf{n}$ are arbitrary, leading to the
following constraint on the space of physical states:
\begin{equation}
(\mathbf{L} + \mathbf{K})|\Psi\rangle = 0\,.
\label{eq:B=1constr}
\end{equation}
The ``grand spin'' $\mathbf{M} = \mathbf{L} + \mathbf{K}$, or its
components, appears quite frequently in what follows. The
states satisfying (\ref{eq:B=1constr}) are linear combinations of the states
$|J,L_3\rangle \otimes |I,K_3\rangle$, whose grand spin is
zero. These are of the form $|J,I;M,M_3\rangle =
|J,J;0,0\rangle$, in the standard notation for adding angular momenta,
where $J=L$ and $I=K$. So the spin $J$ and isospin $I$ have to have
the same magnitude. In addition, $J$ must be half-integral.

The kinetic energy operator is given by
\begin{equation}
\label{T,B=1} T = \frac{1}{2\lambda}\mathbf{J}^2 =
\frac{1}{2\lambda}\mathbf{I}^2\,.
\end{equation}
The eigenvalue of $\mathbf{J}^2$ in states of spin $J$ is $J(J+1)$,
a standard result we will use frequently. Similarly $\mathbf{I}^2$ has eigenvalues
$I(I+1)$. For the lowest energy states, the nucleons with spin/isospin $\half$,
the spin energy in physical units is
\begin{equation}
\frac{1}{2(45.1)}\frac{3}{4}e^3 F_\pi = 21.7\hbox{\,MeV}\,,
\end{equation}
and the total energy is
\begin{equation}
E_{J=1/2,\,I=1/2} = {\cal{M}}_1 + 21.7\hbox{\,MeV} =
986.2\hbox{\,MeV} + 21.7\hbox{\,MeV} = 1008\hbox{\,MeV}\,.
\end{equation}
This is not a bad fit to the nucleon mass $939\hbox{\,MeV}$. The
spin energy is approximately one quarter of its value with the old
parameters, but the higher classical Skyrmion mass makes the
total energy too high. The spin/isospin $\frac{3}{2}$ delta resonances
come out with the too low energy $1095\hbox{\,MeV}$. Of course,
using the old parameter set (\ref{eq:prov}), the nucleon and delta
masses are exactly right.

As the rational map $R(z)=z$ satisfies $-1/\overline{R(z)}=R(-1/\overline{z})$, the parity operator 
$\cal{P}$ is naturally represented by the identity operator. We then find that each of the states described above has positive parity, in agreement with experiment.

\section{$B=2$}
The symmetry of the $B=2$ Skyrmion is $D_{\infty h}$, and the
rational map which approximates this Skyrmion is $R(z)=z^2$. 
The tensors of inertia $U_{ij}$, $V_{ij}$ and $W_{ij}$
are all diagonal, with $U_{11}=U_{22}$, $V_{11}=V_{22}$ and
$W_{11}=W_{22}=0$. We also have that
$U_{33}=\frac{1}{2}W_{33}=\frac{1}{4}V_{33}$, relations which make
the inertia tensor degenerate, a consequence of the axial
symmetry. The degeneracy is resolved by imposing the following
FR constraint on physical states:
\begin{equation}
(L_3+2K_3)|\Psi\rangle = 0\,.
\end{equation}
The discrete symmetry $R(1/z)=1/R(z)$ leads to the FR constraint
\begin{equation}
e^{i\pi L_1}e^{i\pi K_1}|\Psi\rangle = -|\Psi \rangle\,.
\end{equation}
The ground state is then the $J=1$, $I=0$ state $|1,0\rangle
\otimes |0,0\rangle$, which has the quantum numbers of the
deuteron. The first excited state $|0,0\rangle \otimes |
1,0\rangle$ may be identified with the isovector $^1 S_0$ state of
the two-nucleon system.

Using the expressions for the inertia tensors given in
Appendix A, we find that numerically,
\begin{equation}
U_{11}=96.58\,,\, U_{33}=62.94 \hbox{\,\,\,and\,\,\,}
V_{11}=160.61\,.
\end{equation}
The kinetic energy operator is given by \cite{bc}
\begin{equation}
T=\frac{1}{2V_{11}}\mathbf{J}^2 + \frac{1}{2U_{11}}\mathbf{I}^2 -
\left(\frac{1}{2U_{11}} + \frac{2}{V_{11}} -
\frac{1}{W_{33}}\right)K_3^2\,.
\end{equation}
For the ground state, we find (with the conversion factor $e^3 F_\pi$
implied from now on)
\begin{equation}
\label{eq:energy1}
E_{J=1,\,I=0} = {\cal{M}}_2 + 16.3\hbox{\,MeV} =
1949.3\hbox{\,MeV} + 16.3\hbox{\,MeV} = 1966\hbox{\,MeV}\,.
\end{equation}
For the first excited state, we find
\begin{equation}
\label{eq:energy2}
E_{J=0,\,I=1} = {\cal{M}}_2 + 27.1\hbox{\,MeV} =
1976\hbox{\,MeV}\,.
\end{equation}

The experimentally determined mass of the deuteron is 1876\,MeV, with
the proton and neutron constituents only very weakly bound by 2\,MeV.
The $^1 S_0$ state is marginally unbound, with a mass of 1880\,MeV
\cite{bc,isovector}. As our energies (\ref{eq:energy1}) and (\ref{eq:energy2})
exceed the sum of the masses of a proton and a neutron, it would
appear that we have predicted states that are
unbound. However, when we compare $E_{J=1,\,I=0}$ and $E_{J=0,\,I=1}$
to the sum of the masses of two quantized single Skyrmions with spin
$\half$ (calculated in the previous section), these states appear
bound (with binding energies 50\,MeV and 39\,MeV respectively).

While the new parameters are clearly not ideal in the $B=2$
sector, they predict results that are quantitatively
quite accurate. The old parameters more strongly overestimate the binding
energies of these two states \cite{bc}. Also, we calculate the
excitation energy of the $^1 S_0$ state to be 11\,MeV
relative to the deuteron, which is of the correct order of
magnitude, and better than that obtained in \cite{bc} (35\,MeV).

To determine the parities of these two states we observe that the rational map $R(z) = z^2$ has the reflection 
symmetry $-1/\overline{R(z)}=-R(-1/\overline{z})$, and so ${\cal{P}}=e^{i\pi K_3}$. Applying $\cal{P}$
to the allowed states, we find that both have positive parity, in agreement with experiment. 

\section{$B=3$}
The tetrahedrally
symmetric $B=3$ Skyrmion was first quantized in \cite{Carson}. Here we use
the rational map ansatz to simplify the analysis. The Skyrmion is approximated using the map
\begin{equation}
R(z) = \frac{\sqrt{3}iz^2-1}{z^3 - \sqrt{3}iz}\,.
\end{equation}
The symmetry group is generated by two elements. These
correspond to the following symmetries of the rational map:
\begin{eqnarray}
R(-z) &=& -R(z)\,, \\
R\left(\frac{iz+1}{-iz+1}\right)&=&\frac{iR(z)+1}{-iR(z)+1}\,.
\end{eqnarray}
A $\pi$ rotation about the $x_3$-axis in space is equivalent to a
$\pi$ isorotation about the $3$-axis in isospace; and a $2\pi /3$
rotation about the $(x_1+x_2+x_3)$-axis in space is equivalent to
a $2\pi /3$ isorotation about the $(1+2+3)$-axis in isospace. This
leads to the FR constraints
\begin{eqnarray}
\label{eq:b=3_fr1} e^{i\pi L_3}e^{i\pi K_3}|\Psi\rangle &=&
| \Psi\rangle\,, \\
\label{eq:b=3_fr2}
e^{i\frac{2\pi}{3\sqrt{3}}(L_1+L_2+L_3)}
e^{i\frac{2\pi}{3\sqrt{3}}(K_1+K_2+K_3)}|\Psi\rangle &=& |
\Psi\rangle\,.
\end{eqnarray}

There is a spin $\frac{1}{2}$, isospin $\frac{1}{2}$ 
(unnormalised) solution of these constraints,
\begin{equation}
\label{eq:state1} | \Psi \rangle =
\left|\frac{1}{2},\frac{1}{2}\right\rangle \otimes
\left|\frac{1}{2},-\frac{1}{2}\right\rangle -
\left|\frac{1}{2},-\frac{1}{2}\right\rangle \otimes
\left|\frac{1}{2},\frac{1}{2}\right\rangle\,.
\end{equation}
This is the unique state with the same quantum numbers as the hydrogen-3/
helium-3 isodoublet of nuclei in their ground states. The FR constraints
also allow for two distinct states with spin $\frac{3}{2}$ and isospin
$\frac{3}{2}$, given by
\begin{equation}
\label{eq:state2} | \Psi \rangle =
\left|\frac{3}{2},\frac{3}{2}\right\rangle \otimes
\left|\frac{3}{2},-\frac{3}{2}\right\rangle -
\left|\frac{3}{2},\frac{1}{2}\right\rangle \otimes
\left|\frac{3}{2},-\frac{1}{2}\right\rangle +
\left|\frac{3}{2},-\frac{1}{2}\right\rangle \otimes
\left|\frac{3}{2},\frac{1}{2}\right\rangle -
\left|\frac{3}{2},-\frac{3}{2}\right\rangle \otimes
\left|\frac{3}{2},\frac{3}{2}\right\rangle\,,
\end{equation}
and
\begin{equation}
\label{eq:state3} | \Psi \rangle =
\left|\frac{3}{2},-\frac{1}{2}\right\rangle \otimes
\left|\frac{3}{2},-\frac{3}{2}\right\rangle +
\left|\frac{3}{2},-\frac{3}{2}\right\rangle \otimes
\left|\frac{3}{2},-\frac{1}{2}\right\rangle -
\left|\frac{3}{2},\frac{3}{2}\right\rangle \otimes
\left|\frac{3}{2},\frac{1}{2}\right\rangle -
\left|\frac{3}{2},\frac{1}{2}\right\rangle \otimes
\left|\frac{3}{2},\frac{3}{2}\right\rangle\,.
\end{equation}
The first of these has the correct quantum numbers to allow
for its interpretation as a nucleus in which one of the nucleons
is excited to a delta isobar \cite{delta}.

The inertia tensors have been numerically determined for the
rational map given above. They are all diagonal and proportional to
the unit matrix: $U_{ij}=u\delta_{ij}$,
$V_{ij}=v\delta_{ij}$ and $W_{ij}=w\delta_{ij}$. This was to be
expected due to the irreducibility of the action of the
tetrahedral group on ${\mathbb{R}}^3$. Numerically,
\begin{equation}
\label{eq:uvw} u=121.80\,,\, v=418.83 \hbox{\,\,\,and\,\,\,}
w=-80.34\,.
\end{equation}
The kinetic energy operator then takes the following form:
\begin{equation}
T = \frac{1}{2}\frac{1}{uv-w^2}\left[(u-w)\mathbf{J}^2 +
(v-w)\mathbf{I}^2 + w\mathbf{M}^2\right]\,,
\end{equation}
where $\mathbf{M} = \mathbf{L} + \mathbf{K}$. Each of the three
states (\ref{eq:state1},\ref{eq:state2},\ref{eq:state3}) given
above can be rewritten in terms of the basis states
$|J,I;M,M_3\rangle$: the first is proportional to $|\frac{1}{2},\frac{1}{2};0,0\rangle$,
the second to $|\frac{3}{2},\frac{3}{2};0,0\rangle$ and
the third to $|\frac{3}{2},\frac{3}{2};3,2\rangle
- |\frac{3}{2},\frac{3}{2};3,-2\rangle$. They
are thus eigenstates of ${\mathbf{M}}^2$ with eigenvalues 0, 0 and 12
respectively. The energies of the three states are then:
\begin{eqnarray}
E_{J=1/2,\,I=1/2,\,M=0} &=& {\cal{M}}_3 +
\frac{3}{8}\frac{u+v-2w}{uv-w^2} = {\cal{M}}_3
+ 15.4\,\hbox{MeV} = 2895\,\hbox{MeV}\,,\\
E_{J=3/2,\,I=3/2,\,M=0} &=& {\cal{M}}_3 +
\frac{15}{8}\frac{u+v-2w}{uv-w^2} = {\cal{M}}_3
+ 77.1\,\hbox{MeV} = 2957\,\hbox{MeV}\,,\\
E_{J=3/2,\,I=3/2,\,M=3} &=& {\cal{M}}_3 +
\frac{3}{8}\frac{5u+5v+6w}{uv-w^2} = {\cal{M}}_3 +
48.8\,\hbox{MeV} = 2929\,\hbox{MeV}\,.
\end{eqnarray}

These formulae are identical to those obtained in \cite{Carson},
although the numerical values of $u$, $v$ and $w$ are different because of the
rational map approximation. The
average mass of a helium-3 nucleus and a hydrogen-3 nucleus is
2809\,MeV. Our ground state comes to within 4\% of this value. However,
our second state, with an excitation energy of 62\,MeV, is rather too
low in energy to have an ${\rm NN}\Delta$ interpretation, which would require a mass splitting of approximately 300\,MeV with the spin $\frac{1}{2}$
ground state. Using
the old parameter set, one obtains closer agreement with experiment.

To determine the parities of these three states we observe that there is the reflection 
symmetry $-1/\overline{R(iz)}=iR(-1/\overline{z})$, and so ${\cal{P}}=e^{i\frac{\pi}{2}(L_3+K_3)}$. Applying $\cal{P}$
to the allowed states (\ref{eq:state1},\ref{eq:state2},\ref{eq:state3}), we find that they have parities $+$, $+$ and $-$, respectively. We note that the helium-3 and hydrogen-3 ground
states have positive parity. 

\section{$B=4$}
The minimal-energy $B=4$ Skyrmion has $O_h$ symmetry and a cubic shape, and is
described by the rational map
\begin{equation}
\label{eq:b4rm} R(z) = \frac{z^4 + 2\sqrt{3}iz^2 +1}{z^4 -
2\sqrt{3}iz^2 +1}\,.
\end{equation}
This map has the generating symmetries
\begin{eqnarray}
R(iz)&=& 1/R(z)\,, \\
R\left(\frac{iz+1}{-iz+1}\right) &=& e^{i\frac{2\pi}{3}}R(z)\,,
\end{eqnarray}
which lead to the FR constraints
\begin{eqnarray}
e^{i\frac{\pi}{2}L_3} e^{i\pi K_1}|\Psi \rangle &=& |\Psi
\rangle\,, \\
e^{i\frac{2\pi}{3\sqrt{3}}(L_1+L_2+L_3)} e^{i\frac{2\pi}{3}K_3}
| \Psi \rangle &=& |\Psi \rangle\,.
\end{eqnarray}
Seeking simultaneous solutions of these, we
obtain the ground state $|0,0\rangle \otimes |0,0\rangle$.
There exists a spin 2, isospin 1 state given by
\begin{equation}
\label{eq:j2i1}
\left(|2,2\rangle +\sqrt{2}i |2,0\rangle + |2,-2\rangle\right)\otimes |
1,1\rangle - \left(|2,2\rangle -\sqrt{2}i |2,0\rangle + |2,-2\rangle\right)
\otimes |1,-1\rangle\,,
\end{equation}
and a spin 4, isospin 0 state given by \cite{irwin}
\begin{equation}
\left(|4,4\rangle +\sqrt{\frac{14}{5}}|4,0\rangle + |
4,-4\rangle\right)\otimes |0,0\rangle\,.
\end{equation}
The cubic symmetry excludes a spin 2, isospin 0 state.

The tensors of inertia are found to be diagonal, satisfying
$U_{11}=U_{22}$, $V_{ij}=v\delta_{ij}$ and $W_{ij}=0$. Although
the cubic group acts irreducibly on spatial ${\mathbb{R}}^3$, the associated
isospin rotations are reducible, with the ${\mathbb{R}}^3$ of isospace
decomposing into a 2-dimensional and a 1-dimensional subspace. This
is why the inertia tensor $U$ has two independent diagonal
entries, whereas $V$ only has one, and why the cross term $W$
vanishes. Numerically,
\begin{equation}
U_{11}=142.84\,, \, U_{33}=169.41 \hbox{\,\,\,and\,\,\,} v=663.16\,.
\end{equation}
The kinetic energy operator is given by
\begin{equation}
T = \frac{1}{2v}\mathbf{J}^2 + \frac{1}{2U_{11}}\mathbf{I}^2 +
\frac{1}{2}\left(\frac{1}{U_{33}}-\frac{1}{U_{11}}\right)K_3^2\,.
\end{equation}

For the spin 0, isospin 0 ground state, the energy is
simply the static mass of the Skyrmion, ${\cal{M}}_4 =
3679\hbox{\,MeV}$. Comparing this to the mass of the helium-4
nucleus, 3727\,MeV, we see that our prediction comes to within 2\%
of the experimental value. The classical binding energy of the
$B=4$ Skyrmion is significantly larger than that of the $B=3$ or
$B=5$ Skyrmion (see next section). The mean charge radius of the
quantized $B=4$ Skyrmion was calculated using the new parameter
set in \cite{mantonwood} to be 2.13\,fm, which agrees reasonably
well with the experimental value of 1.71\,fm. Walhout \cite{walhout}
calculated this quantity using the old parameter set and taking
into account a number of the vibrational modes, obtaining 1.58\,fm.

For the state (\ref{eq:j2i1}) with spin 2 and isospin 1, the energy is
\begin{equation}
E_{J=2,\,I=1} = {\cal{M}}_4 + 28.7\hbox{\,MeV}\, = 3679.0\hbox{\,MeV} +
28.7\hbox{\,MeV} = 3708\hbox{\,MeV}\,.
\end{equation}
We note here that hydrogen-4, helium-4 and lithium-4 form an
isospin triplet, whose lowest energy state has spin 2, and average
excitation energy 23.7\,MeV relative to the ground state of
helium-4 \cite{energy4}, so here the Skyrmion picture works well.

Finally, for the predicted spin 4, isospin 0 state, we find
\begin{equation}
E_{J=4,\,I=0} = {\cal{M}}_4 + 39.4\hbox{\,MeV}\, = 3679.0\hbox{\,MeV} + 39.4\hbox{\,MeV} = 3718\hbox{\,MeV}\,.
\end{equation}
Such a state of helium-4 has not yet been experimentally observed.
However, predictions for such a state with an excitation energy of
24.6\,MeV have been made \cite{spin4,spin4other}. Our calculation
suggests a slightly larger energy, in the range 30-40\,MeV
(allowing for the discrepancy between our calculation and the data
for the isospin 1 state). The energy levels are summarized in Fig. 3.

\begin{figure}[h!]
\begin{center}
\includegraphics[width=7cm]{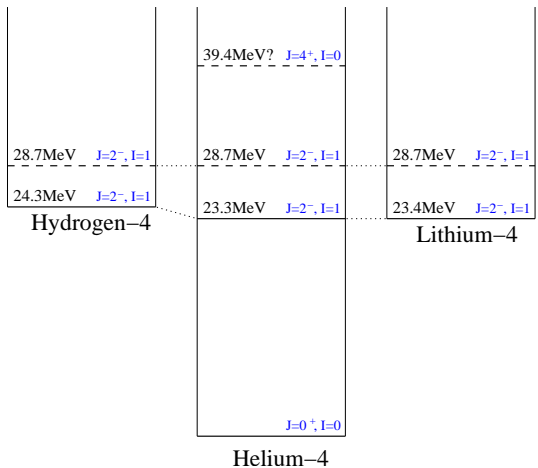}
\label{fig:b4enlvl} \caption{Energy level diagram for the quantized
$B=4$ Skyrmion. Solid lines indicate experimentally observed
states, while dashed lines indicate our predictions.}
\end{center}
\end{figure}

To determine the parities of these three states we observe that the rational map (\ref{eq:b4rm}) has the reflection 
symmetry $-1/\overline{R(z)} = -R(-1/\overline{z})$, and so ${\cal{P}}=e^{i\pi K_3}$. By acting with this 
operator on the physical states, we find that the spin 0, isospin 0 state, and the spin 4, isospin 0 state both have positive parity. On the other hand the spin 2, isospin 1
state has negative parity, and so we find no contradiction with experiment.

\section{$B=5$}
Finding a quantized Skyrmion description of the ground and first
excited states of the helium-5/lithium-5 isodoublet, with spins
$\frac{3}{2}$ and $\frac{1}{2}$, has proved difficult. Physically,
these states are not bound, and they may best be described as a cubic
$B=4$ Skyrmion loosely attracted to a single Skyrmion.

It still remains to determine the symmetries and FR constraints that
might give a lowest energy state of spin $\frac{3}{2}$.
Here we explore in detail the idea floated in \cite{olganick},
that one should consider variants of the rational map, and not
just the one that optimises the classical Skyrmion energy. The
minimal-energy $B=5$ Skyrmion has $D_{2d}$ symmetry, and it can be
approximated by the rational map
\begin{equation}
\label{eq:rmfive}
R(z)=\frac{z(z^4+ibz^2+a)}{az^4+ibz^2+1}\,,\,\,\,a=-3.07\,, \,\,
b=3.94 \,.
\end{equation}
The ground state obtained from this map \cite{irwin} has spin
$\frac{1}{2}$ and isospin $\frac{1}{2}$, which is inconsistent
with the observed spin $\frac{3}{2}$ ground states of helium-5 and
lithium-5. The Skyrmion has this shape up to a pion mass $m\simeq
1$. However, if higher $m$ is considered, the symmetry of the Skyrmion
might change, and the above ground state become unstable.
Therefore more symmetric solutions which apparently have higher energy
might be the ones describing the true Skyrmion, and are worth investigating.

When $b=0$ the rational map (\ref{eq:rmfive}) has $D_{4h}$
symmetry, and it acquires octahedral symmetry when in addition
$a=-5$. Octahedral symmetry has been previously considered
\cite{krusch} and it leads to a ground state with spin
$\frac{5}{2}$ and isospin $\frac{1}{2}$. Let us therefore consider
the $D_{4h}$-symmetric map (which could in fact be restricted to
$C_4$ symmetry)
\begin{equation}
\label{eq:b5rm}
R(z) = \frac{z(z^4 + a)}{az^4+1}\,, \quad a\neq-5\,.
\end{equation}
This map has the generating symmetries
\begin{eqnarray}
R(iz) &=& iR(z)\,,\\
R(1/z) &=& 1/R(z)\,,
\end{eqnarray}
which lead to the FR constraints
\begin{eqnarray}
\label{eq:b5fr1}
e^{i\frac{\pi}{2}L_3}e^{i\frac{\pi}{2}K_3}|\Psi\rangle &=& - |
\Psi\rangle\,, \\e^{i\pi L_1}e^{i\pi K_1} |\Psi\rangle &=&
|\Psi\rangle\,. \label{eq:b5fr2}
\end{eqnarray}
Seeking simultaneous solutions, we obtain a ground state with
$J=\frac{3}{2}$ and isospin $I=\frac{1}{2}$, given by
\begin{equation}
| \Psi\rangle=\left|\frac{3}{2},\frac{3}{2}\right\rangle\otimes
\left|\frac{1}{2},\frac{1}{2}\right\rangle +
\left|\frac{3}{2},-\frac{3}{2}\right\rangle\otimes
\left|\frac{1}{2},-\frac{1}{2}\right\rangle\,.
\end{equation}
This is the spin we are looking for. There are two excited states with
$J=\frac{5}{2}$ and $I=\frac{1}{2}$, most easily written in terms
of $|J, I; M,M_3\rangle$, where
$\mathbf{M}=\mathbf{L}+\mathbf{K}$:
\begin{equation}
\label{eq:b5fe}
| \Psi\rangle = 
\left( \left|\frac{5}{2},
 \frac{1}{2};\, 3,2\right
\rangle -
\left|\frac{5}{2},
 \frac{1}{2};\, 3,-2\right
\rangle \right)
+ c_{\pm}\left(\left|\frac{5}{2},
 \frac{1}{2};\, 2,2\right
\rangle + \left|\frac{5}{2},
 \frac{1}{2};\, 2,-2\right
\rangle \right)\,,
\end{equation}
with $c_{\pm}$ evaluated in Appendix C.
The FR constraints allow for a further excited state
with $J=\frac{1}{2}$ and $I = \frac{3}{2}$.

$D_4$ symmetry implies that the tensors of inertia
are diagonal, with $U_{11}=U_{22}$, $V_{11}=V_{22}$
and $W_{11}=W_{22}$, which leads to the expression for the kinetic
energy operator
\begin{equation}
T=\frac{1}{2}\Bigg\{\frac{1}{(U_{11}V_{11}-W_{11}^2)}\left[
U_{11}(\mathbf{J}^2-L_3^2) + V_{11}(\mathbf{I}^2-K_3^2) +
W_{11}(\mathbf{M}^2-\mathbf{J}^2-\mathbf{I}^2-2L_3K_3)\right]
\end{equation}
\begin{equation*}
+\frac{1}{(U_{33}V_{33}-W_{33}^2)}\left[ U_{33}L_3^2 + V_{33}K_3^2
+ 2W_{33}L_3K_3\right] \Bigg\}\,. \label{eq:b5ke}
\end{equation*}
The energy of the ground state is therefore
\begin{equation}
E_{J=3/2,\,I=1/2}={\cal{M}}_5 +
\frac{3U_{11}+V_{11}}{4(U_{11}V_{11}-W_{11}^2)}
+\frac{9U_{33}+V_{33}+6W_{33}}{8(U_{33}V_{33}-W_{33}^2)}\,.
\end{equation}
The numerical value of the energy depends
on the parameter $a$ in the rational map, which has yet to be
determined.

We now argue that the $D_{4h}$ symmetry which we are considering
is justified even if octahedral symmetry ($a=-5$) provides us with a
slightly lower classical energy. The dependence of the classical
energy on $a$ is shown in Fig. 4, whereas
the quantum energy is a strictly increasing function of $a$ near
$a=-5$ (see Fig. 5). Therefore the total energy
achieves its minimum just below $a=-5$ (see Fig. 6),
the quantum energy being much smaller than the classical one. Taking
$a=-5.0025$ we find that
\begin{eqnarray}
U_{11}=203.41\,, \quad U_{33}&=&203.36\,, \quad V_{11}=1333.49\,,
\quad V_{33}=1332.96\,,\\ \nonumber W_{11}&=&-186.54\,, \quad
 W_{33}=-186.61\,.
\end{eqnarray}
The static Skyrmion mass ${\cal{M}}_5$, for this value of $a$, is calculated to be
5101\,MeV. The energies of the four states given above are
then:
\begin{eqnarray}
E_{J=3/2,\,I=1/2}&=&{\cal{M}}_5 + 8.2\hbox{\,MeV} = 5109\hbox{\,MeV}\,,\\
E_{J=5/2,\,I=1/2,\, c_{-}}&=&{\cal{M}}_5 + 12.5\hbox{\,MeV}
=5114\hbox{\,MeV}\,,
\\ E_{J=5/2,\,I=1/2, \, c_{+}}&=&{\cal{M}}_5 + 12.9\hbox{\,MeV}
=5114
\hbox{\,MeV}\,,\\
E_{J=1/2,\,I=3/2}&=&{\cal{M}}_5 + 26.9\hbox{\,MeV} =5128
\hbox{\,MeV}\,.
\end{eqnarray}
So, the achievement of the correct spin
$\frac{3}{2}$ for the ground state comes at a price.
Firstly, the slightly excited $J=\frac{1}{2}$ state is not allowed by the
FR constraints. Secondly, by comparing $E_{J=3/2,\,I=1/2}$ to the average mass of the helium-5
and lithium-5 nuclei (4668\,MeV), we see that our prediction is
some 10\% from the experimental value, whereas for the
$D_{2d}$-symmetric Skyrmion there was an almost exact match to the
helium-5/lithium-5 ground state energy.

The $D_{4h}$-symmetric map (\ref{eq:b5rm}) satisfies $-1/\overline{R(z)} = R(-1/\overline{z})$. 
The parity operator could therefore be represented by the identity operator. However, we may also choose ${\cal{P}}=e^{2\pi i \mathbf{n} \cdot \mathbf{L}}$, where 
$\mathbf{n}$ is any unit vector. If we make this choice, then
each of the states described above has negative parity. We note that the ground states of helium-5 and lithium-5 have negative parities. 

\begin{figure}[h!]
\begin{center}
\includegraphics[width=7.5cm]{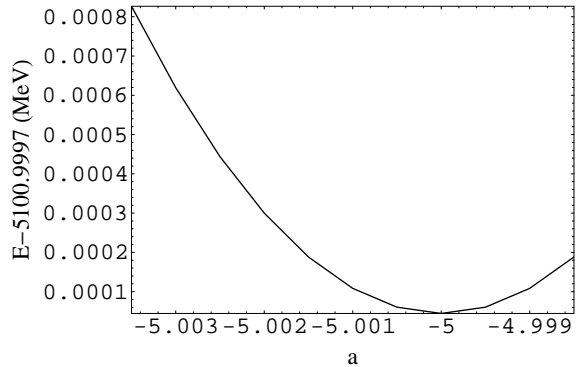}
\caption{Classical energy of the $B=5$ Skyrmion as a function of
$a$.}
\end{center}
\end{figure}

\begin{figure}[h!]
\begin{center}
\includegraphics[width=7.5cm]{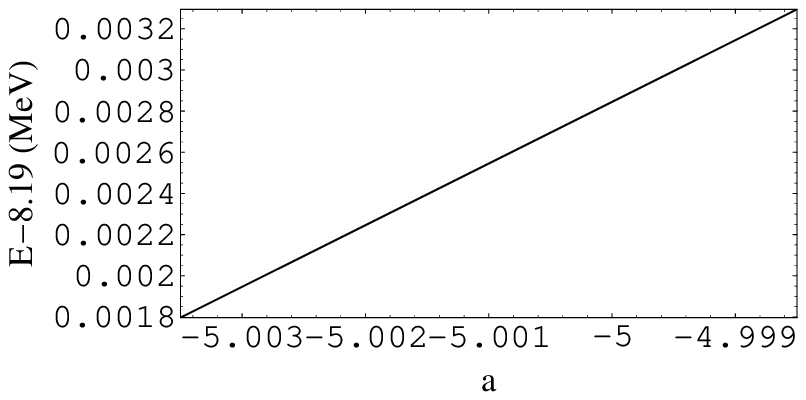}
\caption{Quantum energy of the $B=5$ Skyrmion as a function of
$a$.}
\end{center}
\end{figure}

\begin{figure}[h!]
\begin{center}
\includegraphics[width=7.5cm]{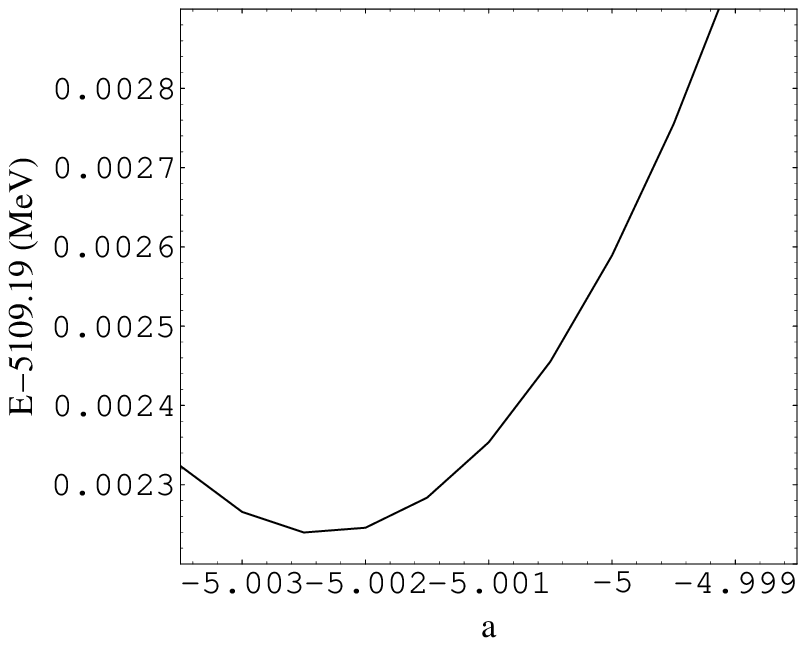}
\caption{Total energy of the $B=5$ Skyrmion as a function of $a$.}
\end{center}
\end{figure}

\section{$B=6$}
The minimal-energy $B=6$ Skyrmion has $D_{4d}$ symmetry, and is
well-approximated using the rational map
\begin{equation}
R(z)=\frac{z^4 + ia }{ z^2 (iaz^4 + 1)}\,.
\end{equation}
This map has the generating symmetries
\begin{eqnarray}
R(iz)&=&-R(z)\,, \\
R(1/z)&=&1/R(z)\,,
\end{eqnarray}
which lead to the FR constraints
\begin{eqnarray}
e^{i\frac{\pi}{2} L_3}e^{i\pi K_3}|\Psi\rangle &=&
|\Psi\rangle \,, \\
e^{i\pi L_1}e^{i\pi K_1}|\Psi\rangle &=& -|\Psi\rangle \,.
\end{eqnarray}
These constraints allow for the existence of states $|1,0\rangle
\otimes |0,0\rangle$, $|3,0\rangle \otimes |0,0\rangle$,
$|0,0\rangle \otimes |1,0\rangle$, $|2,0\rangle \otimes
|1,0\rangle$ and $|5,0\rangle \otimes |0,0\rangle$.

A numerical search over the parameter $a$ in the rational map
shows that the integral $\cal{I}$ is minimized at $a = 0.16$.
However, it was suggested in \cite{mantonwood} that allowing a
slight deformation of the rational map would lead to more accurate
predictions. In particular, it was found that by setting $a =
0.1933$, and using the new parameter set, one obtains a quantum
quadrupole moment in agreement with experiment. In what follows,
we set $a = 0.1933$.

The inertia tensors have been computed for this
rational map, and are each found to be diagonal, satisfying
$U_{11}=U_{22}$, $V_{11}=V_{22}$ and $W_{11}=W_{22}=0$.
Numerically,
\begin{equation}
U_{11}=215.84 \,, \, U_{33}=230.77 \,, \, V_{11}=1525.99 \,, \,
V_{33}=1493.66 \hbox{\,\,\,and\,\,\,} W_{33}=-105.45 \,.
\end{equation}
The kinetic energy operator is given by:
\begin{equation}
T=\frac{1}{2V_{11}}\left[\mathbf{J}^2-L_3^2\right] +
\frac{1}{2U_{11}}\left[\mathbf{I}^2-K_3^2\right] +
\frac{1}{2(U_{33}V_{33}-W_{33}^2)}\left[U_{33}L_3^2 + V_{33}K_3^2
+ 2W_{33}L_3K_3\right]\,.
\end{equation}
The static Skyrmion mass, ${\cal{M}}_6$, is calculated to be
5601\,MeV, which is precisely equal to the mass of the lithium-6
nucleus (the new parameter set was determined such that this would
be the case -- in \cite{mantonwood} we estimated the spin energy
for spin 1, isospin 0 to be approximately 1\,MeV, and then
neglected this small quantity). The energy eigenvalues
corresponding to the five states given above are then:
\begin{eqnarray}
E_{J=1,\,I=0} &=& {\cal{M}}_6 + \frac{1}{V_{11}} = {\cal{M}}_6
+ 1.7\,\hbox{MeV} = 5602\,\hbox{MeV}\,,\\
E_{J=3,\,I=0} &=& {\cal{M}}_6 + \frac{6}{V_{11}} = {\cal{M}}_6
+ 10.3\,\hbox{MeV} = 5611\,\hbox{MeV}\,,\\
E_{J=0,\,I=1} &=& {\cal{M}}_6 + \frac{1}{U_{11}} = {\cal{M}}_6
+ 12.1\,\hbox{MeV} = 5613\,\hbox{MeV}\,,\\
E_{J=2,\,I=1} &=& {\cal{M}}_6 + \frac{1}{U_{11}} +
\frac{3}{V_{11}} = {\cal{M}}_6 + 17.2\,\hbox{MeV} =
5618\,\hbox{MeV}\,,\\
E_{J=5,\,I=0} &=& {\cal{M}}_6 + \frac{15}{V_{11}} = {\cal{M}}_6 +
25.7\,\hbox{MeV} = 5626\,\hbox{MeV}\,.
\end{eqnarray}
We may identify these with isospin 0 states of lithium-6, and with
states of the helium-6, lithium-6 and beryllium-6 nuclei, which
together form an isospin triplet (see Fig. 7).
The assumption in \cite{mantonwood} that the spin kinetic energy
of the state $|1,0\rangle \otimes |0,0\rangle$ is of order 1\,MeV is
clearly justified.

\begin{figure}[h!]
\begin{center}
\includegraphics[width=7.5cm]{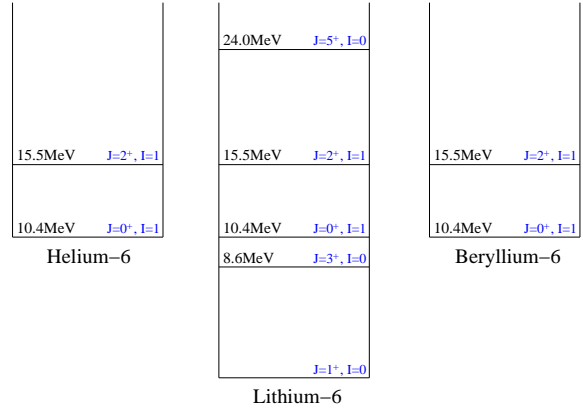}
\caption{Energy level diagram for the quantized $B=6$
  Skyrmion. Energies are given relative to the spin 1, isospin 0 ground state.}
\end{center}
\end{figure}

\begin{figure}[h!]
\begin{center}
\includegraphics[width=7.5cm]{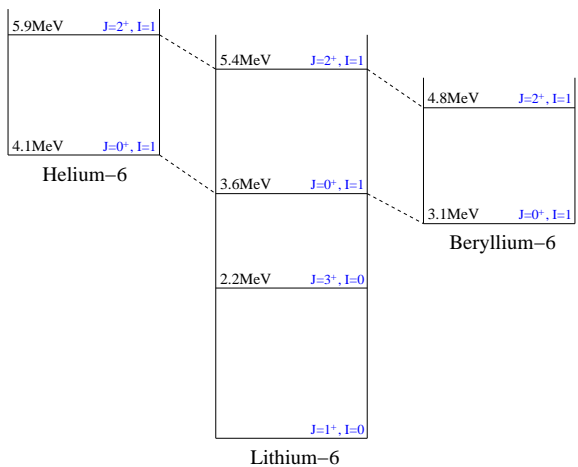}
\label{fig:b6exp} \caption{Energy level diagram for nuclei with $B=6$.}
\end{center}
\end{figure}

Spin and isospin excitation energies relative to the lithium-6 ground
state are experimentally known for these nuclei
\cite{energy567} (see Fig. 8). The ground state of the lithium-6 nucleus is
identified with the state $|1,0\rangle \otimes |0,0\rangle$, and
there is an excited state $|3,0\rangle \otimes |0,0\rangle$, with
excitation energy $2.2\,\hbox{MeV}$. Lithium-6 has a further spin
0 excited state with excitation energy $3.6\,\hbox{MeV}$; this is
joined by the lowest energy states of the helium-6 and beryllium-6
nuclei, which relative to the ground state of lithium-6 have energies of
$4.1\,\hbox{MeV}$ and $3.1\,\hbox{MeV}$, respectively,
to form the spin 0 isotriplet $|0,0\rangle \otimes |1,0\rangle$.
Similarly $|2,0\rangle \otimes |1,0\rangle$ is identified with
the spin 2 excited states of the isotriplet. A spin 5 excited state of lithium-6 has not been seen
experimentally. We, however, predict the existence of such a state
with excitation energy higher than those of the other states so
far discussed.

The splitting between the various spin and isospin states of the
Skyrmion is clearly too large; the predicted quantum energies are
roughly four times the experimental values. This may be connected
to the fact that lithium-6 is an odd-odd nucleus. However, we have performed
the same calculation using the old parameter set, and have found
that this gives even wider gaps between the energy levels. The
new parameter set is therefore not perfect, but is certainly an
improvement. Furthermore, the ratios of the relative excitation
energies given by
\begin{equation}
\frac{E_{J=0,\,I=1} - E_{J=1,\,I=0}}{E_{J=3,\,I=0} -
E_{J=1,\,I=0}} = 1.2
\end{equation}
and
\begin{equation}
\frac{E_{J=2,\,I=1} - E_{J=1,\,I=0}}{E_{J=0,\,I=1} -
E_{J=1,\,I=0}} = 1.5
\end{equation}
correspond well to experimental data for these nuclei, for which
the first ratio is $3.6/2.2 = 1.6$ and the second is $5.4/3.6 =
1.5$.

To determine the parities of the states given above we firstly 
observe the reflection symmetry 
\begin{equation}
-1/\overline{R(e^{i\frac{\pi}{4}}z)} = -iR(-1/\overline{z})\,.
\end{equation}
The parity operator can therefore be represented as ${\cal{P}} = e^{i\frac{\pi}{4}L_3}e^{-i\frac{\pi}{2}K_3}$. If we make this choice, then each of the states given above has positive parity, in agreement with experiment. 

\section{$B=7$}
\label{sec:7} Here, as for $B=5$, quantizing the Skyrmion of
lowest energy gives states with the wrong spins to match the
nuclear data. The minimal-energy $B=7$ Skyrmion has icosahedral
symmetry, and is described by the rational map
\begin{equation}
R(z)=\frac{7z^5+1}{z^2(z^5-7)}\,.
\end{equation}
This map leads to a ground state with $J=\frac{7}{2}$,
$I=\frac{1}{2}$, a spin which appears experimentally as the second
excited state of the lithium-7/beryllium-7 isospin doublet.
Experimentally, the ground state has spin $\frac{3}{2}$.

There are many ways in which the icosahedral symmetry might be
broken, allowing for the appearance of a $J=\frac{3}{2}$,
$I=\frac{1}{2}$ state in the spectrum. The most interesting
possibility, in our opinion, is the breaking of the $C_3$
symmetry, while preserving $D_5$ symmetry. This leads to a ground
state with $J=\frac{3}{2}$ and $I=\frac{1}{2}$. So let us consider the $D_5$-symmetric map
\begin{equation}
\label{eq:b7rm}
R(z)=\frac{az^5+1}{z^2(z^5-a)}\,, \quad a\neq 7\,,
\end{equation}
where $a = 7$ restores the icosahedral symmetry. The
generating symmetries of this map are
\begin{eqnarray}
R(e^{i\frac{2\pi}{5}}z)&=&e^{-i\frac{4\pi}{5}}R(z)\,,\\
R(-1/z)&=&-1/R(z)\,,
\end{eqnarray}
which lead to the FR constraints
\begin{eqnarray}
\label{eq:frb7} e^{i
\frac{2\pi}{5}L_3}e^{-i\frac{4\pi}{5}K_3}\left|\Psi\right\rangle
&=&-\left|\Psi\right\rangle\,, \\
e^{i\pi L_2}e^{i\pi
K_2}\left|\Psi\right\rangle&=&-\left|\Psi\right\rangle\,.
\end{eqnarray}
The ground state with $J =\frac{3}{2}$ and $I =\frac{1}{2}$ is
\begin{equation}
\left| \Psi\right\rangle=\left|\frac{3}{2},
\frac{3}{2}\right\rangle\otimes\left|\frac{1}{2},-\frac{1}{2}\right\rangle
+\left|\frac{3}{2},
-\frac{3}{2}\right\rangle\otimes\left|\frac{1}{2},\frac{1}{2}\right\rangle
\,,
\end{equation}
the first excited state with $J=\frac{5}{2}$ and $I=\frac{1}{2}$ is
\begin{equation}
\left| \Psi\right\rangle=\left|\frac{5}{2},
\frac{3}{2}\right\rangle\otimes\left|\frac{1}{2},-\frac{1}{2}\right\rangle
-\left|\frac{5}{2},
-\frac{3}{2}\right\rangle\otimes\left|\frac{1}{2},\frac{1}{2}\right\rangle
\,,
\end{equation}
and there exist two further excited states with $J=\frac{7}{2}$,
$I=\frac{1}{2}$, given by
\begin{eqnarray}
\left| \Psi^1\right\rangle&=&\left|\frac{7}{2},
\frac{3}{2}\right\rangle\otimes\left|\frac{1}{2},-\frac{1}{2}\right\rangle
+\left|\frac{7}{2},
-\frac{3}{2}\right\rangle\otimes\left|\frac{1}{2},\frac{1}{2}\right\rangle
\,, \\\left| \Psi^2\right\rangle&=&\left|\frac{7}{2},
\frac{7}{2}\right\rangle\otimes\left|\frac{1}{2},\frac{1}{2}\right\rangle
-\left|\frac{7}{2},
-\frac{7}{2}\right\rangle\otimes\left|\frac{1}{2},-\frac{1}{2}\right\rangle
\,.
\end{eqnarray}
States with $I=\frac{3}{2}$ are also allowed. In particular, there
is one spin $\frac{1}{2}$ state:
\begin{equation}
\left| \Psi\right\rangle = \left|\frac{1}{2},
\frac{1}{2}\right\rangle\otimes\left|\frac{3}{2},\frac{3}{2}\right\rangle
-\left|\frac{1}{2},
-\frac{1}{2}\right\rangle\otimes\left|\frac{3}{2},-\frac{3}{2}\right\rangle
\,,
\end{equation}
and two spin $\frac{3}{2}$ states:
\begin{eqnarray}
\left| \Psi^1\right\rangle&=&\left|\frac{3}{2},
\frac{3}{2}\right\rangle\otimes\left|\frac{3}{2},-\frac{1}{2}\right\rangle
-\left|\frac{3}{2},
-\frac{3}{2}\right\rangle\otimes\left|\frac{3}{2},\frac{1}{2}\right\rangle
\,, \\\left| \Psi^2\right\rangle&=&\left|\frac{3}{2},
\frac{1}{2}\right\rangle\otimes\left|\frac{3}{2},\frac{3}{2}\right\rangle
+\left|\frac{3}{2},
-\frac{1}{2}\right\rangle\otimes\left|\frac{3}{2},-\frac{3}{2}\right\rangle
\,.
\end{eqnarray}

The inertia tensors are found to be diagonal, with
$U_{11}=U_{22}$, $V_{11}=V_{22}$ and $W_{11}=W_{22}=0$, leading to
the kinetic energy operator:
\begin{equation}
\label{eq:b7energy}
T=\frac{1}{2V_{11}}\left[\mathbf{J}^2-L_3^2\right] +
\frac{1}{2U_{11}}\left[\mathbf{I}^2-K_3^2\right] +
\frac{1}{2(U_{33}V_{33}-W_{33}^2)}\left[U_{33}L_3^2 + V_{33}K_3^2
+ 2W_{33}L_3K_3\right]\,.
\end{equation}
The energy of the ground state is given by
\begin{equation}
\label{eq:energyseven} E_{J=3/2,\,I=1/2}= {\cal{M}}_7 +
\frac{3}{4V_{11}}+\frac{1}{4U_{11}}+\frac{9U_{33}+V_{33}-6W_{33}}{8(U_{33}V_{33}-W_{33}^2)}\,.
\end{equation}
The static Skyrmion mass, ${\cal{M}}_7$, is found to be close to
6328\,MeV. The dependence of the classical and quantum energies on
$a$  are shown in Fig. 9 and Fig. 10 respectively. Looking for the
value of $a$ giving the minimum of the total energy, we obtain
$a=7.002$ (see Fig. 11). For this value of $a$, we find numerically
\begin{equation}
U_{11}=246.27\,, \,\, U_{33}=246.26\,, \,\, V_{11}=1873.03\,, \,\,
V_{33}=1872.76\,, \,\, W_{33}=0.04\,,
\end{equation}
and
\begin{equation}
E_{J=3/2,\,I=1/2}={\cal{M}}_7 +6.6\hbox{\,MeV} =
6335\hbox{\,MeV}\,,
\end{equation}
to be compared to the average mass of the lithium-7 and
beryllium-7 nuclei which is 6534\,MeV.

\begin{figure}[h!]
\begin{center}

\includegraphics[width=7.5cm]{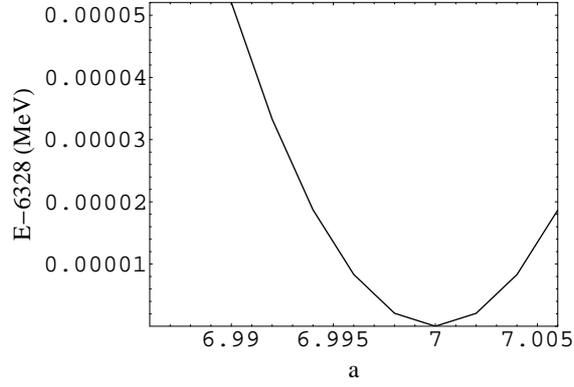}
\caption{Classical energy of the $B=7$ Skyrmion as a function of
$a$.}

\end{center}
\end{figure}

\begin{figure}[h!]
\begin{center}

\includegraphics[width=7.5cm]{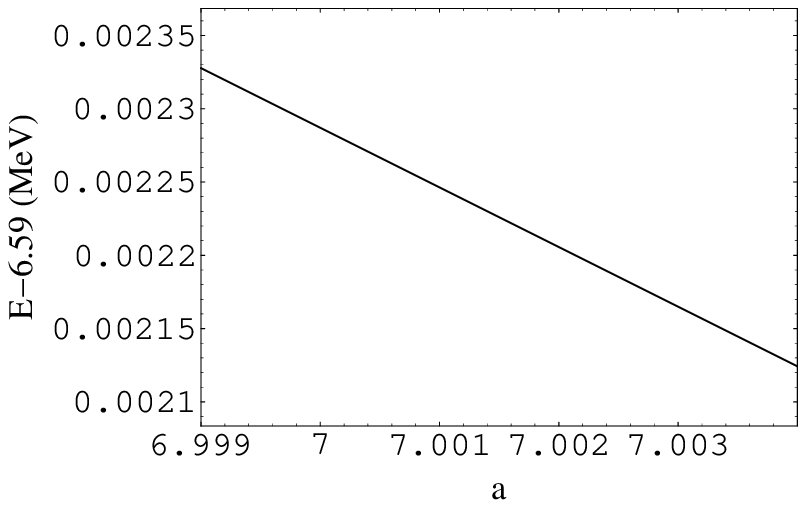}
\caption{Quantum energy of the $B=7$ Skyrmion as a function of
$a$.}

\end{center}
\end{figure}

\begin{figure}[h!]
\begin{center}

\includegraphics[width=7.5cm]{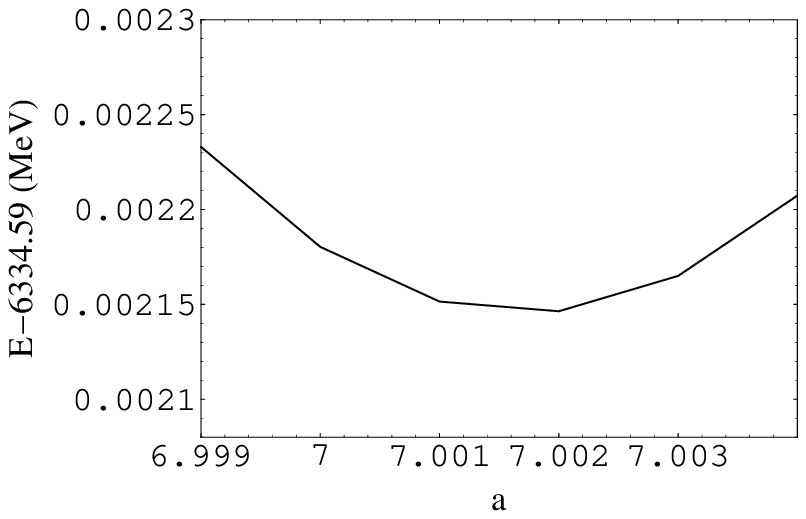}
\caption{Total energy of the $B=7$ Skyrmion as a function of $a$.}
\end{center}
\end{figure}

For the excited states we find that:
\begin{eqnarray}
E_{J = 5/2, \, I=1/ 2}&=& {\cal{M}}_7 + 10.1\hbox{\,MeV} =
6338\hbox{\,MeV}\,, \\
E^1_{J = 7/2, \, I=1/ 2}&=&{\cal{M}}_7 +
15.0\hbox{\,MeV} = 6343\hbox{\,MeV}\,, \\
E^2_{J = 7/2, \, I=1/ 2}&=&{\cal{M}}_7 +
15.0\hbox{\,MeV} = 6343\hbox{\,MeV}\,, \\
E_{J = 1/2, \, I=3/ 2}&=&{\cal{M}}_7 + 20.4\hbox{\,MeV} =
6348\hbox{\,MeV}\,, \\
E^1_{J = 3/2, \, I=3/ 2}&=&{\cal{M}}_7 +
22.5\hbox{\,MeV} =
6351\hbox{\,MeV}\,, \\
E^2_{J = 3/2, \, I=3/ 2}&=&{\cal{M}}_7 +
22.5\hbox{\,MeV} = 6351 \hbox{\,MeV}\,.
\end{eqnarray}
There are two main problems with the above spectrum. One is
the absence of the $J=\frac{1}{2}$, $I=\frac{1}{2}$ state, and the
other is the appearance of the $J=\frac{5}{2}$, $I=\frac{1}{2}$ state as the first
excitation. We could try to overcome this problem by noticing that
the first two excited states in the experimental lithium-7 and
beryllium-7 spectra are in a sense \emph{anomalous}: they have
very low excitation energy, and the spin-energy correspondence is
reversed. As was discussed in the introduction it is possible that
such excitations cannot be described by our usual approach, and we
need to allow for some vibrational modes or consider a Skyrmion
of a different shape. Possibly, the states we find above could correspond
to the ones lying above the lowest energy $J=\frac{7}{2}$, $I=\frac{1}{2}$ excited state.
This interpretation fits rather well to the experimental
data. The second problem
is more difficult to tackle within this framework. The
value of $a$ being very close to 7 leads to a configuration which is
nearly $C_3$--symmetric. This fact is reflected in the spectrum:
we have two spin $\frac{7}{2}$ and two isospin $\frac{3}{2}$
states whose energies are almost indistinguishably close. This is
not reflected in the experimental data. Let us therefore consider
a smaller $a$ and see if there is a better fit to the spectrum. Another
advantage of this approach is that it helps to partially
overcome the first problem as well. Indeed, by looking through a large range of $a$ we find that at $a=2$ the energies of
the states given above are, in increasing order,
\begin{eqnarray}
E_{J = 3/2, \, I=1/ 2}&=&{\cal{M}}_7 + 6.3\hbox{\,MeV}, \\
E^2_{J = 7/2, \, I=1/ 2}&=&{\cal{M}}_7 + 9.3\hbox{\,MeV}, \\
E_{J = 5/2, \,I=1/ 2}&=&{\cal{M}}_7 + 9.4\hbox{\,MeV}, \\
E^1_{J = 7/2, \, I=1/ 2}&=&{\cal{M}}_7 + 13.7\hbox{\,MeV}, \\
E_{J = 1/2, \, I=3/ 2}&=&{\cal{M}}_7 + 19.7\hbox{\,MeV}, \\
E^1_{J = 3/2, \, I=3/ 2}&=&{\cal{M}}_7 + 19.9\hbox{\,MeV}, \\ 
E^2_{J = 3/2, \, I=3/2}&=&{\cal{M}}_7 + 21.6\hbox{\,MeV}.
\end{eqnarray}
The energy of the $J=\frac{5}{2}$, $I=\frac{1}{2}$
state is now higher than the energy of one of the $J=\frac{7}{2}$,
$I=\frac{1}{2}$ states, and lower than that of the other, in
agreement with experiment. This achievement, however,
comes at a price as the classical energy is now some $10\%$
higher than the experimental value. Figs. 12 and 13 are energy level diagrams for the quantized $D_5$-symmetric $B=7$ 
Skyrmion, with $a=2$, and for the $B=7$ nuclei, respectively.

The symmetry breaking we have just considered is only one of
the ways in which icosahedral symmetry might be broken. It is
possible that a different breaking has to be
considered in order to better understand the spectrum of the excited
states, in particular the low-lying $J=\frac{1}{2}$, $I=\frac{1}{2}$ state. 
It is also possible that with the increase of the pion
mass the configuration will eventually break up into a $B=4$ and
a $B=3$ part, which is suggested by the very low energy for break up of lithium-7
into helium-4 plus a triton.
\begin{figure}[h!]
\begin{center}
\includegraphics[width=8cm]{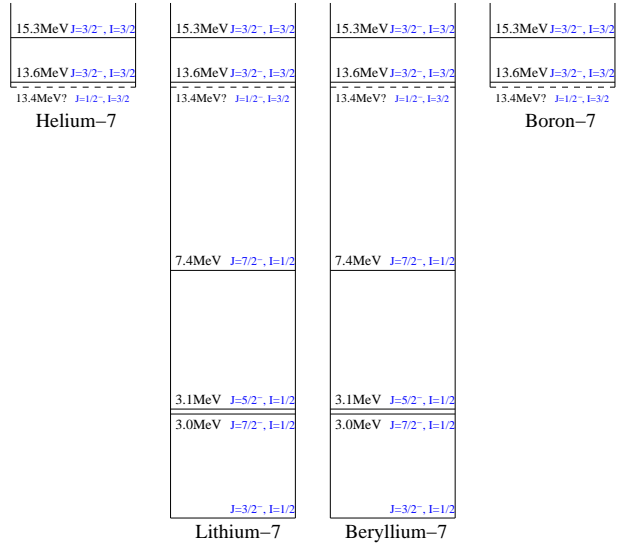}
\caption{Energy level diagram for the quantized $B=7$ Skyrmion. A putative
$J=\frac{1}{2}^{-}$ isoquartet is represented by dashed lines.}
\end{center}
\end{figure}

\begin{figure}[h!]
\begin{center}
\includegraphics[width=8cm]{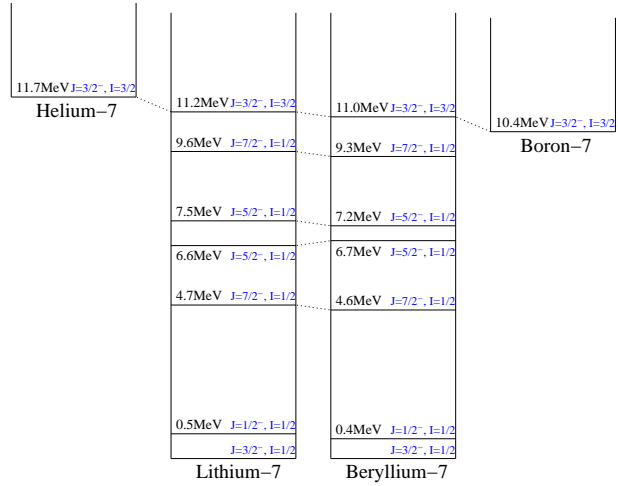}
\label{fig:b7exp} \caption{Energy level diagram for nuclei with
$B=7$.}
\end{center}
\end{figure}

The $D_{5}$-symmetric map (\ref{eq:b7rm}) satisfies $-1/\overline{R(z)} = R(-1/\overline{z})$. As for $B=5$, we find it favourable to choose 
${\cal{P}}=e^{2\pi i \mathbf{n} \cdot \mathbf{L}}$, where 
$\mathbf{n}$ is any unit vector. If we make this choice, then
each of the states described above has negative parity, in agreement with experiment.

\section{$B=8$}
In this section we introduce some new ideas for estimating the
moments of inertia of the $B=8$ Skyrmion, and hence the excitation
energies of the quantum states. It is believed that for our new
parameter set, the minimal-energy classical solution resembles two
touching $B=4$ cubes (see Fig. 14) \cite{bms}. Here the rational
map ansatz is not a good approximation, so our previous methods of
calculation are no longer valid. Despite this, it is convenient to
note that a field which is qualitatively of the right form, with
the correct symmetries, can be obtained from a rational map, and
this enables one to determine the allowed spin/isospin/parity
states. There are also classical Skyrmion
solutions which are well approximated by the rational map ansatz,
and have only very slightly greater energy than the double cube.
We consider these first.

For pion mass parameter between 0 and approximately 1, the minimal-energy 
$B=8$ Skyrmion has $D_{6d}$ symmetry, and is well-approximated 
by the rational map
\begin{equation}
R(z)=\frac{z^6-ia}{z^2(iaz^6-1)}\,,  \,\,\, a=0.14\,,
\end{equation}
which has the symmetries
\begin{eqnarray}
R\left(e^{i\frac{\pi}{3}}z\right)&=&e^{-i\frac{2\pi}{3}}R(z)\,, \\
R(1/z)&=&1/R(z)\,,
\end{eqnarray}
leading to the FR constraints
\begin{eqnarray}
e^{i\frac{\pi}{3} L_3}e^{-i\frac{2 \pi}{3} K_3}|\Psi \rangle &=& |
\Psi \rangle\,, \\
e^{i\pi L_1}e^{i\pi K_1}|\Psi \rangle &=& |\Psi \rangle\,.
\end{eqnarray}
The ground state is then determined to be
$|0,0\rangle\otimes|0,0\rangle$, and the first excited state is
$|2, 0\rangle \otimes |0, 0\rangle$, in agreement with states of
the beryllium-8 nucleus. However, this Skyrmion becomes unstable
once the pion mass parameter exceeds 1. The true minimum is then
described by two $B=4$ cubes placed together, and as a first
approximation to this in terms of a rational map we consider the
$O_h$-symmetric map (whose Wronskian vanishes on the 14 faces of a
truncated octahedron):
\begin{equation}
\label{eq:rmeight} R(z)=\frac{z^8+4\sqrt{3}z^6-10z^4+4\sqrt{3}z^2+1}{z^8-4\sqrt{3}z^6-10z^4-4\sqrt{3}z^2+1}\,,
\end{equation}
whose symmetries
\begin{eqnarray}
R(iz)&=&1/R(z)\,, \\
R\left(\frac{iz+1}{-iz+1}\right) &=& \frac{-\sqrt{3} + R(z)}{1 +
\sqrt{3}\,R(z)}\,,
\end{eqnarray}
lead to the FR constraints
\begin{eqnarray}
e^{i\frac{\pi}{2}L_3}e^{i\pi K_1}|\Psi\rangle &=& |\Psi\rangle\,, \\
e^{i\frac{2\pi}{3\sqrt{3}}(L_1+L_2+L_3)}e^{-i\frac{2\pi}{3}K_2}|\Psi\rangle
&=& |\Psi\rangle\,.
\end{eqnarray}
Here the ground state is again $|0, 0\rangle \otimes|0, 0\rangle$,
but the $|2, 0\rangle \otimes |0, 0\rangle$ state is not allowed. The inertia tensors 
for this rational map are found to be diagonal, satisfying $U_{11}=U_{33}$, $V_{11}=V_{22}=V_{33}$ and $W_{ij}=0$.

However, the $O_h$ symmetry is too strong for the description of
two cubes, and has to be relaxed to $D_{4h}$ symmetry.
Therefore we consider next
\begin{equation}
\label{eq:rmeight1} R(z)=\frac{z^8+bz^6-az^4+bz^2+1}{z^8-bz^6-az^4-bz^2+1}\,,
\end{equation}
where $a=10$ and $b=4\sqrt{3}$ restores the $O_h$
symmetry. The rational map ansatz then gives a better
approximation to the double cube Skyrmion, but only slightly
because, for example, $U=-1$ at the origin with the rational
map ansatz, whereas for the true solution, $U=-1$ at points near
the cube centres. However, it has the right symmetry, and is good
enough to determine the allowed spin/isospin states. The FR
constraints are now
\begin{eqnarray}
\label{eq:b8fr1}
e^{i\frac{\pi}{2}L_3}e^{i\pi K_1}|\Psi\rangle &=& |\Psi\rangle \,, \\
\label{eq:b8fr2} e^{i\pi L_1}|\Psi\rangle &=& |\Psi\rangle \,,
\end{eqnarray}
which again allows a $|2, 0\rangle \otimes |0, 0\rangle$ state. The
inertia tensors have the same symmetry properties as for the
octahedral map, with the exceptions that $U_{11} \neq U_{33}$ and
$V_{33} \neq V_{11} = V_{22}$. This leads to the kinetic energy
operator:
\begin{equation}
\label{eq:firstT}
T =\frac{1}{2V_{11}}\left[\mathbf{J}^2 - L_3^2\right] +
\frac{L_3^2}{2V_{33}} + \frac{K_1^2}{2U_{11}} + \frac{K_2^2}{2U_{22}} + \frac{K_3^2}{2U_{33}}\,.
\end{equation}
To determine the parities of states, we observe the 
rational map (\ref{eq:rmeight1}) has the 
reflection symmetry $-1/\overline{R(z)}=-1/R(-1/\overline{z})$. The parity operator can 
therefore be represented by ${\cal{P}}=e^{i\pi K_2}$. 

To progress, we now work directly with two cubic
$B=4$ Skyrmions separated along the $x_3$-axis, and find the moments of inertia of the resulting
structure using the parallel axis theorem (ignoring the
interaction of the cubes). The top cube is rotated by $\frac{\pi}{4}$ about the $x_3$-axis 
relative to the standard orientation corresponding to (\ref{eq:b4rm}). The bottom
cube is rotated by $-\frac{\pi}{4}$ about the $x_3$-axis 
relative to the standard orientation. One difficulty here is in determining the
separation of the cubes. The picture in Fig. 14 suggests that the
separation is the value of $r$ where the profile function becomes
close to zero. From Fig. 2 we see that it is reasonable to take
$r= 1.8$ leading to the separation in question being
$d=r/\sqrt{3}=1.04$ in dimensionless units. Then
\begin{eqnarray}
V^{(B=8)}_{11}&=&V^{(B=8)}_{22}=2V^{(B=4)}_{11} + {\cal{M}}d^2=2706\,, \\
V^{(B=8)}_{33}&=&2V^{(B=4)}_{33}=1326 \,.
\end{eqnarray}
where ${\cal{M}}=1277$ (in dimensionless units) is the classical mass of two $B=4$ Skyrmions.
The isospin moments of inertia are simply given by
\begin{eqnarray}
U^{(B=8)}_{11}&=&U^{(B=8)}_{22}=2U^{(B=4)}_{11}=286 \,, \\
U^{(B=8)}_{33}&=&2U^{(B=4)}_{33}=339 \,.
\end{eqnarray}
The equality of $U_{11}$ and $U_{22}$, which we do not expect to be exactly satisfied by the true $B=8$ 
solution, simplifies (\ref{eq:firstT}) to 
\begin{equation}
T =\frac{1}{2V_{11}}\left[\mathbf{J}^2 - L_3^2\right] +
\frac{1}{2U_{11}}\left[\mathbf{I}^2 - K_3^2\right] +
\frac{L_3^2}{2V_{33}} + \frac{K_3^2}{2U_{33}}\,.
\end{equation}
The ground state has quantum energy zero, so its total energy is simply the classical Skyrmion mass.
The additional quantum energy of the spin 2, isospin 0 state is
2.9\,MeV, which is a very good match
to the experimental value of 3\,MeV \cite{energy8910}. There are a lot of further
excited states, consistent with the FR constraints (\ref{eq:b8fr1}) and (\ref{eq:b8fr2}),
whose wavefunctions, energies and parities are
presented in Table 1. Figs. 15 and 16 are energy level diagrams for the quantized $B=8$ Skyrmion, and for the $B=8$ nuclei, respectively. We see a good agreement with experiment for positive parity states, and the appearance of some
negative parity states which have not yet been observed
experimentally. Of particular interest is the appearance of the
$J=0$, $I=1$ negative parity state. If found, it could be a new
ground state of the lithium-8 nucleus. The detection of the latter
might be very difficult experimentally. We have
also found quintets of $I=2$ states. The lowest of these, with spin 0, have
been detected experimentally with excitation energies very close
to our predictions, and include the helium-8 and carbon-8 ground states.

It remains worthwhile to find the optimal values of $a$ and $b$ in the
rational map (\ref{eq:rmeight1}). Ideally, the classical
mass should not be very far away from the experimental mass of
the beryllium-8 ground state which is 7455\,MeV, and the
moments of inertia should be
comparable with the ones we get from the double cube approach.
The second condition is more difficult to achieve since the
rational map is defined on a sphere, and cannot
exactly reproduce a double cube configuration. Looking through a
range of possible $a$ and $b$ values we find that the optimal map
is given approximately by
\begin{equation}
\label{eq:rmeight2} 
R(z)=\frac{z^8+\frac{13\sqrt{3}}{2}z^6-20z^4+\frac{13\sqrt{3}}{2}z^2+1}{z^8-\frac{13\sqrt{3}}{2}z^6-20z^4-\frac{13\sqrt{3}}{2}z^2+1}\,,
\end{equation}
leading to the following moments of inertia
\begin{eqnarray}
V_{11}=V_{22}&=&2901\,, \\
V_{33}&=&2214\,,\\
U_{11}&=&308\,,\\
U_{22}&=&268\,,\\
U_{33}&=&283\,.
\end{eqnarray}
We have recalculated the energies of the states in Table 1, using the kinetic energy operator (\ref{eq:firstT}) and
formulae for the energy levels of an asymmetrical top \cite{landau}. The energy of the first excited state is
\begin{equation}
E_{J=2,\,I=0}={\cal{M}}_8 + \frac{3}{V_{11}} = 7529\hbox{\,MeV} +
2.7\hbox{\,MeV} = 7531\hbox{\,MeV}\,,
\end{equation}
which is only slightly worse than the double cube approach.
However, for further excited states the discrepancy in results
increases, making the advantages of the double cube approach more
evident.

\begin{figure}[h!]
\begin{center}
\includegraphics{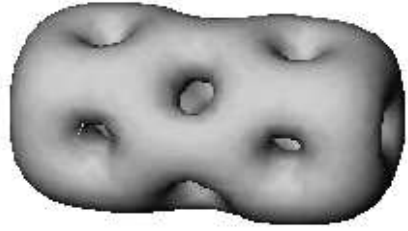}
\label{fig:b4} \caption{Baryon density isosurface for the numerically relaxed $B=8$ Skyrmion 
with $m \approx 1$, resembling two touching $B=4$ Skyrmions.}
\end{center}
\end{figure}

\begin{table}[h!]
\begin{center}
\vspace{0.2cm}
\begin{tabular}{|c|c|c|c|c|}
\hline
{$\textbf{J},\,\textbf{I}$}&\textbf{Wavefunction}&\textbf{Parity}&$\textbf{\it{E}}_{dc}$\,\textbf{(MeV)}&$\textbf{\it{E}}_{rm}$\,\textbf{(MeV)} \\
\hline
$0,0$& $|0,0\rangle\otimes|0,0\rangle$ &$+$&0&0 \\
\hline
$2,0$& $|2,0\rangle\otimes|0,0\rangle$ &$+$&2.9&2.7 \\
\hline
\multirow{2}{*}{$4,\,0$}& $|4,0\rangle\otimes|0,0\rangle$ &$+$&9.7&9.0 \\
&$(|4,4\rangle+|4,-4\rangle)\otimes|0,0\rangle$ &$+$&17.7&11.2 \\
\hline
$0,1$& $|0,0\rangle\otimes(|1,1\rangle - |1,-1\rangle )$ &$-$&8.4&9.5 \\
\hline
\multirow{3}{*}{$2,1$}& $|2,0\rangle\otimes(|1,1\rangle - |1,-1\rangle$ &$-$&11.3&12.2 \\
&$(|2,2\rangle+|2,-2\rangle)\otimes(|1,1\rangle+|1,-1\rangle)$&$+$&13.3&12.1 \\
&$(|2,2\rangle+|2,-2\rangle)\otimes|1,0\rangle$&$-$&14.1&12.4 \\
\hline 
\multirow{2}{*}{$3,\,1$}&$(|3,2\rangle-|3,-2\rangle)\otimes(|1,1\rangle +|1,-1\rangle)$&$+$&16.2&14.8 \\
&$(|3,2\rangle-|3,-2\rangle)\otimes|1,0\rangle$&$-$&16.9&15.1\\
\hline
\multirow{4}{*}{$4,1$}& $|4,0\rangle\otimes(|1,1\rangle-|1,-1\rangle)$&$-$&18.1&18.5 \\
&$(|4,2\rangle+|4,-2\rangle)\otimes(|1,1\rangle +|1,-1\rangle)$&$+$&20.1&18.4 \\
&$(|4,2\rangle+|4,-2\rangle)\otimes|1,0\rangle$&$-$&20.8&18.7 \\
&$(|4,4\rangle+|4,-4\rangle)\otimes(|1,1\rangle-|1,-1\rangle)$&$-$&26.1&20.7 \\
\hline
\multirow{3}{*}{$0,2$}& $|0,0\rangle\otimes(|2,2\rangle + |2,-2\rangle)$&$+$&24.6&26.3 \\
&$|0,0\rangle\otimes(|2,1\rangle+|2,-1\rangle)$&$-$&26.7&26.4 \\
&$|0,0\rangle\otimes|2,0\rangle$&$+$&27.4&28.6 \\
\hline
\multirow{5}{*}{$2,2$}& $|2,0\rangle\otimes(|2,2\rangle+|2,-2\rangle)$&$+$&27.5&29.0 \\
&$(|2,2\rangle+|2,-2\rangle)\otimes(|2,2\rangle-|2,-2\rangle)$&$-$&29.5&30.8 \\
&$|2,0\rangle\otimes (|2,1\rangle+|2,-1\rangle)$&$-$&29.6&29.1\\
&$|2,0\rangle\otimes|2,0\rangle$&$+$&30.3&31.3 \\
&$(|2,2\rangle+|2,-2\rangle)\otimes(|2,1\rangle-|2,-1\rangle)$&$+$&31.6&31.6 \\
\hline
\end{tabular}
\caption{Energies and parities of the $B=8$  allowed states. $E_{dc}$ and $E_{rm}$ are the quantum energies obtained using the double cube approach and rational map ansatz respectively.}
\end{center}
\end{table}

\begin{figure}[h!]
\begin{center}
\includegraphics[width=10cm]{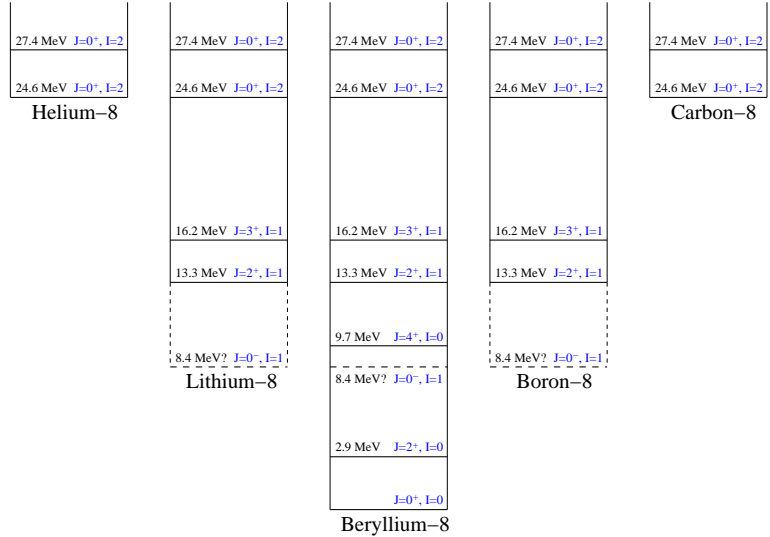}
\label{fig:b8theor} \caption{Energy level diagram for the quantized $B=8$ Skyrmion, using the double cube approach. A putative $J=0^-$ isotriplet is represented by dashed lines.}
\end{center}
\end{figure}

\begin{figure}[h!]
\begin{center}
\includegraphics[width=10cm]{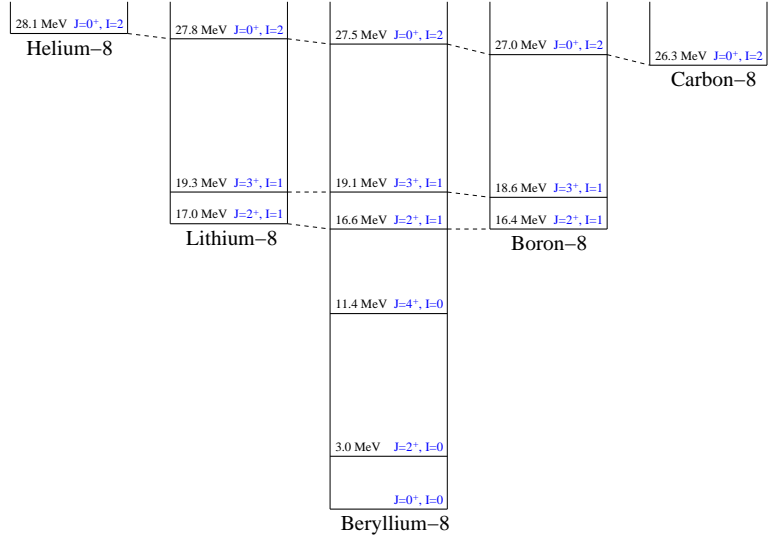}
\label{fig:b8exp} \caption{Energy level diagram for nuclei with $B=8$.}
\end{center}
\end{figure}

\section{Conclusion}
\label{sec:conc}
The rational map ansatz simplifies the classification of the allowed spin and 
isospin states of quantized Skyrmions, and has enabled us to estimate their
moments of inertia and energy spectra. The results are promising, and provide support
for the interpretation of Skyrmions as nuclei. We have obtained the
correct spin, parity and isospin quantum numbers for the ground states and
various excited states in most cases, and the quantum energies of
excited states are reasonably close to the experimental values. We
have also been able to predict some excited
states that have not yet been observed. The new parameter set for the Skyrme model, with
which we have been working throughout, has provided better results
than the traditional parameter set for the larger values of $B$. We
have also put into effect a new approach for some Skyrmions of odd
baryon number, in particular for $B=7$. By deforming the highly symmetric minimal-energy Skyrmion,
we have been able to reproduce the spins of the experimental ground state and several excited states.
We have given the first estimates of the energies of
quantum states based on the double cube $B=8$ Skyrmion, and
similar methods should be applicable to the multi-cube solutions
for $B=12, 16$ and beyond, presented in \cite{bms}.

The calculations presented here are subject to a number of
limitations. Firstly, we consider the semiclassical quantization,
in which only the collective coordinates for rotations and isospin
rotations are considered. A more accurate procedure would have to
take into account further degrees of freedom, which we refer to as
vibrational modes. Our current understanding is that the Skyrme model
provides a description of nuclear physics in which nucleons are partially merged, and their orientations
in space and isospace are highly correlated. In a sense, this is the opposite of a naive shell model, in which nucleons
move in a potential,
and are to first approximation uncorrelated. A more realistic model would possibly
lie somewhere between these two extremes.
Allowing the individual Skyrmions, or subclusters of Skyrmions, to move relative to each other, and
performing a quantization of these degrees of freedom,
would be a significant refinement to our approach. In so doing, some missing
low-lying experimentally observed states of
nuclei may appear.
These include
the low-lying excited states with $J=\frac{1}{2}$ and $I=\frac{1}{2}$
that are present for $B=5$ and $B=7$.

The work here should be taken further by working with the exact
Skyrmion solutions, and not just the rational map approximation to
these solutions. Classical energies and moments of inertia will
change, though we hope not drastically. Further investigation of
the effect of varying the dimensionless pion mass parameter is
also warranted. The length scale of the Skyrmions is quite
sensitive to this. Possibly, an increased parameter will create an
instability in the Skyrmions with $B=5$ or $B=7$, thereby
justifying our arguments for changing the symmetries.  

\subsection*{Acknowledgments}
We would like to thank Steffen Krusch for helpful discussions. SWW
thanks Bernard Piette for providing the C++ code with which the
rational map profile functions were numerically determined. OVM
would like to thank EPSRC for the award of a Dorothy Hodgkin
Scholarship, and SWW would like to thank PPARC for a research
studentship. NSM carried out part of this work while visiting the 
Institute for Advanced Study, Princeton, and thanks the Institute and its Director
for hospitality.

\appendix
\section{Appendix: Inertia Tensors}
The tensors of inertia for rational map Skyrmions may be expressed
in the form:
\begin{equation}
\Sigma_{ij} = 2 \int \sin^2 f
\,\frac{C_{\Sigma_{ij}}}{(1+|R|^2)^2} \left(1+f'^2+\frac{\sin^2
f}{r^2} \left(\frac{1+|z|^2}{1+|R|^2}\left|\frac{dR}{dz}\right|
\right)^2 \right) d^3 x\,,
\end{equation}
where $\Sigma = (U,V,W)$ and the quantities $C_{U_{ij}}$ are given
by
\begin{eqnarray}
C_{U_{11}} &=& |1-R^2|^2\,, \\
C_{U_{22}} &=& |1+R^2|^2\,, \\
C_{U_{33}} &=& 4|R|^2\,, \\
C_{U_{12}} = C_{U_{21}} &=& -2 \,\Im R^2\,, \\
C_{U_{13}} = C_{U_{31}} &=& 2\,(|R|^2 -1)\, \Re R\,, \\
C_{U_{23}} = C_{U_{32}} &=& 2\,(|R|^2 -1)\, \Im R\,,
\end{eqnarray}
the quantities $C_{V_{ij}}$ are given by
\begin{eqnarray}
C_{V_{11}} &=& |1-z^2|^2 \left|\frac{dR}{dz}\right|^2\,, \\
C_{V_{22}} &=& |1+z^2|^2 \left|\frac{dR}{dz}\right|^2\,,\\
C_{V_{33}} &=& 4|z|^2 \left|\frac{dR}{dz}\right|^2\,, \\
C_{V_{12}} = C_{V_{21}} &=& -2\, \Im z^2 \,\left|\frac{dR}{dz}\right|^2\,, \\
C_{V_{13}} = C_{V_{31}} &=& 2\, \Re \left(|z|^2 z - \bar{z}
\right) \,
\left|\frac{dR}{dz}\right|^2\,, \\
C_{V_{23}} = C_{V_{32}} &=& 2\, \Im \left(|z|^2 z + \bar{z}
\right) \, \left|\frac{dR}{dz}\right|^2\,,
\end{eqnarray}
and finally, the quantities $C_{W_{ij}}$ are given by
\begin{eqnarray}
C_{W_{11}} &=& \Re \left((1-z^2)(1-\bar{R}^2)\frac{dR}{dz}\right)\,, \\
C_{W_{22}} &=& \Re \left((1+z^2)(1+\bar{R}^2)\frac{dR}{dz}\right)\,, \\
C_{W_{33}} &=& 4\, \Re \left(\bar{R}z\frac{dR}{dz}\right)\,, \\
C_{W_{12}} &=& -\Im \left((1+z^2)(1-\bar{R}^2)\frac{dR}{dz}\right)\,, \\
C_{W_{13}} &=& -2\, \Re \left(z(1-\bar{R}^2)\frac{dR}{dz}\right)\,, \\
C_{W_{23}} &=& -2\, \Im \left(z(1+\bar{R}^2)\frac{dR}{dz}\right)\,, \\
C_{W_{21}} &=& \Im \left((1-z^2)(1+\bar{R}^2)\frac{dR}{dz}\right)\,,\\
C_{W_{31}} &=& -2\, \Re \left(\bar{R}(1-z^2)\frac{dR}{dz}\right)\,, \\
C_{W_{32}} &=& 2\, \Im \left(\bar{R}(1+z^2)\frac{dR}{dz}\right)\,.
\end{eqnarray}

\section{Appendix: Old Parameters}
Here we collect some data on moments of inertia, in Skyrme units,
calculated with the dimensionless pion mass parameter $m=0.528$
that emerges from the calibration of \cite{an}. The following
results are novel, as they were obtained using the rational map
ansatz and the formulae in Appendix A, and extend from $B=1$ up to $B=4$. For $B=1$ the rational
map ansatz is exact, so our result should agree with that of
\cite{an}, and indeed it does. For $B=2,3$ our results can be
compared with the moments of inertia calculated from the exact
Skyrmion solutions (with the same $m$) as given by
\cite{bc,Carson}. This allows us to investigate the
accuracy of the rational map ansatz for these Skyrmions.

The notation is as in the Sections 6 to 9 above. For $B=1$
\begin{equation}
\lambda = 62.85\,.
\end{equation}

For $B=2$
\begin{equation}
U_{11}=135.43\,,\,\, U_{33}=86.59 \hbox{\,\,\,and\,\,\,}
V_{11}=221.88\,.
\end{equation}
Comparing these numbers to those obtained in \cite{bc} using the
exact numerical solution ($U_{11}=127.8$, $U_{33}=86.9$ and
$V_{11}=200.2$), we see that the rational map ansatz has enabled
us to obtain quite accurate moments of inertia. We recall that the
old parameter set led to a model of the deuteron which was much
too tightly bound.

For $B=3$
\begin{equation}
u=170.01\,,\,\, v=576.09 \hbox{\,\,\,and\,\,\,} w=-109.47\,.
\end{equation}
These were evaluated in \cite{Carson}, using the exact numerical
solution ($u=136$, $v=435$ and $w=-91$).

For $B=4$
\begin{equation}
U_{11}=197.60\,,\,\, U_{33}=236.49 \hbox{\,\,\,and\,\,\,} v=911.45\,.
\end{equation}
These numbers were calculated using the same procedure that we
have used throughout, but with the old parameters. Walhout
\cite{walhout} performed a different style of analysis for the
$B=4$ Skyrmion, and unfortunately we are unable to directly
compare our results for the individual components of the inertia
tensors.

\section{Appendix: Coefficients of Wavefunctions}
The FR constraints do not determine all coefficients in
the wavefunctions. Usually finding these constants is trivial, but
in some cases (as in the $B=5$ first and second excited states) one has to be
more careful. As an illustration let us consider the constants $c_{\pm}$
in (\ref{eq:b5fe}). The solutions of (\ref{eq:b5fr1},\ref{eq:b5fr2}) form a subspace of Hilbert space,
which is transformed into itself when acted upon by the operator of the kinetic energy (\ref{eq:b5ke}).
Therefore, the eigenvectors of the operator will define the
wavefunctions we are looking for. In terms of the moments of
inertia, $c_{\pm}$ is given by
\begin{equation}
c_{\pm}=\frac{b_2+5b_1-a_1-5a_2\pm\sqrt{(b_2+5b_1-a_1-5a_2)^2
+20(a_1-a_2)(b_1-b_2)}} {2\sqrt{5}(b_1-b_2)}
\end{equation}
where
\begin{eqnarray}
a_1&=&\frac{1}{8} \left(\frac{10U_{11}+2V_{11}+20W_{11}}{U_{11}V_{11}-W_{11}^2}
+\frac{25U_{33}+V_{33}-10W_{33}}{U_{33}V_{33}-W_{33}^2}\right)\,, \\
a_2&=&\frac{1}{8} \left(\frac{26U_{11}+2V_{11}+4W_{11}}{U_{11}V_{11}-W_{11}^2}
+\frac{9U_{33}+V_{33}+6W_{33}}{U_{33}V_{33}-W_{33}^2}\right)\,, \\
b_1&=&\frac{1}{8} \left(\frac{10U_{11}+2V_{11}-4W_{11}}{U_{11}V_{11}-W_{11}^2}
+\frac{25U_{33}+V_{33}-10W_{33}}{U_{33}V_{33}-W_{33}^2}\right)\,, \\
b_2&=&-\frac{1}{8}\left(\frac{26U_{11}+2V_{11}-20W_{11}}{U_{11}V_{11}-W_{11}^2}
+\frac{9U_{33}+V_{33}+6W_{33}}{U_{33}V_{33}-W_{33}^2}\right)\,.
\end{eqnarray}
The quantum energy of these states is given by
\begin{equation}
E=\frac{a_1+5a_2+c_{\pm}\sqrt{5}(b_1-b_2)}{6}\,.
\end{equation}

\end{document}